\documentclass[nofootinbib,twocolumn,superscriptaddress,showpacs,prd,aps,amsmath,amssymb]{revtex4}
\usepackage{graphicx}
\usepackage{color}
\usepackage{grffile}
\usepackage{amssymb}
\usepackage{mathtools}
\usepackage{bm} 
\usepackage{mathrsfs} 
\usepackage[normalem]{ulem}

\newcommand{\Madm}{M_{\rm ADM}}
\newcommand{\MK}{M_{\rm K}}
\newcommand{\MP}{M_{\rm P}}

\newcommand{\Msol}{M_\odot}

\newcommand{\beq}{\begin{equation}} 
\newcommand{\eeq}{\end{equation}} 
\newcommand{\beqn}{\begin{eqnarray}} 
\newcommand{\eeqn}{\end{eqnarray}} 
\newcommand{\pa}{\partial}
\newcommand{\na}{\nabla}

\newcommand{\gabd}{g_{\alpha\beta}}

\newcommand{\gmabd}{\gamma_{ab}}

\newcommand{\tgmabd}{\tilde\gamma_{ab}}

\newcommand{\habd}{h_{ab}}

\newcommand{\fabd}{f_{ab}}

\newcommand{\tbeta}{\tilde{\beta}}

\newcommand{\albe}{{\alpha\beta}}

\newcommand{\Tabd}{T_{\alpha\beta}}

\newcommand{\Gabd}{G_{\alpha\beta}}

\newcommand{\Fabu}{F^{\alpha\beta}}
\newcommand{\Fabd}{F_{\alpha\beta}}

\newcommand{\QE}{Q^{\rm E}}

\newcommand{\zD}{{\raise1.0ex\hbox{${}^{\ \circ}$}}\!\!\!\!\!D}
\newcommand{\alone}{{\raise0.5ex\hbox{${}^{\ 1}$}}\!\!\!\!\alpha}

\newcommand{\dSa}{dS_{\alpha}}

\newcommand{\dl}{\delta}

\newcommand{\Dl}{\Delta}

\newcommand{\Lie}{\mbox{\pounds}}

\newcommand{\nalam}{\mathrel{\raise0.9ex\hbox{$^\lambda$}\mkern-14mu
\lower0.0ex\hbox{$\nabla$}}}

\newcommand{\Nrf}{{N_r^{\rm f}}}
\newcommand{\Nrm}{{N_r^{\rm m}}}

\newcommand{\zeroD}{{\raise1.0ex\hbox{${}^{\ \circ}$}}\!\!\!\!\!D}

\newcommand{\zLap}{{\raise1.0ex\hbox{${}^{\ \circ}$}}\!\!\!\!\Delta}
\newcommand{\zna}{{\raise1.0ex\hbox{${}^{\ \circ}$}}\!\!\!\!\!\nabla}
\newcommand{\zS}{{\raise1.0ex\hbox{${}^{\ \circ}$}}\!\!\!\!\!S}

\newcommand{\cocal}{{\sc cocal}}

\newcommand{\Bpolmax}{{B^{\rm max}_{\rm pol}}}
\newcommand{\Btormax}{{B^{\rm max}_{\rm tor}}}

\newcommand{\Bpol}{{B_{\rm pol}}}
\newcommand{\Btor}{{B_{\rm tor}}}

\newcommand{\rhoc}{\rho_{\rm c}}
\newcommand{\prhoc}{(p/\rho)_{\rm c}}
\newcommand{\presc}{p_{\rm c}}
\newcommand{\Omegac}{\Omega_{\rm c}}

\newcommand{\ToverW}{{\cal T}/|{\cal W}|}
\newcommand{\MoverW}{{\cal M}/|{\cal W}|}
\newcommand{\PioverW}{{\Pi}/|{\cal W}|}
\newcommand{\Ivir}{I_{\rm vir}}
\newcommand{\IviroverW}{I_{\rm vir}/|{\cal W}|}

\begin{document}
\title{Magnetically supramassive and hypermassive compact stars}

\author{K\=oji Ury\=u}
\email{koji.uryu@gmail.com}
\affiliation{
Department of Physics, University of the Ryukyus, Senbaru 1, 
Nishihara, Okinawa 903-0213, Japan}
\author{Shijun Yoshida}
\email{yoshida@astr.tohoku.ac.jp}
\affiliation{
Astronomical Institute, Tohoku University, Aramaki-Aoba, Aoba, Sendai 980-8578, Japan}
\author{Eric Gourgoulhon}
\email{eric.gourgoulhon@obspm.fr}
\affiliation{LUX, Observatoire de Paris, Universit\'e PSL, CNRS, \\ Sorbonne Universit\'e, 92190 Meudon, France}
\affiliation{Laboratoire de Math\'ematiques de Bretagne Atlantique, CNRS UMR 6205, 
Universit\'e de Bretagne Occidentale, 6 avenue Victor Le Gorgeu, 29200 Brest, France}
\author{Charalampos Markakis}
\email{markakis@illinois.edu}
\affiliation{Department of Applied Mathematics \& Statistics, 
Stony Brook University, Stony Brook, NY 11794, USA}
\affiliation{National Center for Supercomputing Applications, 
University of Illinois at Urbana-Champaign, Urbana, IL 61801, USA}
\author{Kotaro Fujisawa}
\email{kotaro.fujisawa@gmail.com}
\affiliation{
Department of Liberal Arts, Kogakuin University, 
Nakano-cho 2665-1, Hachioji, Tokyo 192-0015, Japan}
\author{Antonios Tsokaros}
\email{tsokaros@illinois.edu}
\affiliation{
Department of Physics, University of Illinois Urbana-Champaign, Urbana, IL 61801, USA}
\affiliation{National Center for Supercomputing Applications, 
University of Illinois Urbana-Champaign, Urbana, IL 61801, USA}
\affiliation{Research Center for Astronomy and Applied Mathematics, 
Academy of Athens, Athens 11527, Greece}
\author{Keisuke Taniguchi}
\email{ktngc@cs.u-ryukyu.ac.jp}
\affiliation{
Department of Physics, University of the Ryukyus, Senbaru 1, 
Nishihara, Okinawa 903-0213, Japan}
\author{Mina Zamani}
\email{m_zamani@znu.ac.ir}
\affiliation{
Department of Physics, University of Zanjan, 
P.O.~Box 45195-313, Zanjan, Iran}
\author{Lambros Boukas}
\email{lboukas@gmail.com}
\affiliation{LAB Scientific Computing Inc., Miami, Florida 33176, USA}
%

%

\date{\today}  

\begin{abstract} 
It is known that the mass of a magnetized relativistic compact star 
is larger than that of non-magnetized one for the same equation of state 
and central density, albeit the excess of mass is sizable only if 
the magnetic fields are strong enough, $B\sim 10^{17}$--$10^{18}$G.  
Using our recently developed numerical code \cocal, 
we systematically compute such compact star solutions in equilibrium 
associated with mixed poloidal and toroidal magnetic fields, 
and show the magnetically supramassive solutions whose masses exceed by more than 
10\% the maximum mass of the static and spherically symmetric solutions.  
For some extremely strong magnetic field configurations, we also obtain 
solutions more massive than the maximum mass of the uniformly rotating solutions 
at the Kepler (mass-shedding) limit, namely magnetically hypermassive solutions.  
\end{abstract} 


\maketitle
 
\section{Introduction}

For a certain equation of state (EOS) of a high density matter, 
the mass $M$ of a sequence of solutions of the Tolman-Oppenheimer-Volkoff (TOV) 
equation (a sequence of hydrostatic equilibriums in relativistic gravity in 
static and spherically symmetric spacetime) 
parametrized by the central density, the $M(\rhoc)$ curve, exhibits a maximum 
(see, e.g. \cite{TOV,Shapiro:1983du}).  
Arguments of a turning point theorem as well as a linear stability analysis 
prove that the maximum mass is related to the point of radial instability.  
Since the maximum mass differs depending on the EOS models, observations 
of massive neutron stars may rule out the EOS models which cannot support 
the observed mass (e.g. \cite{NuclearEOS}).  

On top of this classic argument for the static and spherically symmetric 
relativistic stars (TOV solutions hereafter), 
it is known that the stationary and axisymmetric rapidly rotating relativistic stars 
may have masses higher than the maximum mass of the TOV solutions (for the same EOS).  
For the case with uniform rotation \cite{CST,CST94}, the excess of the maximum mass 
compared to that of the TOV solutions becomes around $\sim15-20\%$, 
and it can be much higher for the cases with differential rotations \cite{RNSdiff}.  
The latter differentially rotating solutions are considered as models for 
the proto neutron stars or the binary neutron star merger remnants \cite{RNSdiff2}.  
Such uniformly rotating compact stars with mass higher than the maximum mass 
of TOV solutions are called supramassive, and the differentially rotating stars 
with mass higher than the maximum mass of the uniformly rotating supramassive 
solution may be also called hypermassive.  

It has been discussed that the maximum mass can be increased substantially 
also because of extremely strong magnetic fields (see, e.g.~\cite{Suvorov:2021ymy}).  
Therefore, in principle, one can obtain \emph{magnetically supramassive} or 
\emph{magnetically hypermassive} solutions, whose excess of mass relative to the corresponding 
unmagnetized solution is due exclusively to the magnetic fields instead of rotation.  
As such configurations are largely deformed due to the magnetic fields, numerical methods 
are the most effective approaches to study such solutions.  Such magnetically 
supramassive as well as hypermassive solutions were demonstrated in the very first 
numerical computations of relativistic stars associated with extremely strong 
purely poloidal magnetic field \cite{Bocquet:1995je}, closely studied in 
\cite{Cardall:2000bs}, and also in several calculations under 
different setups (see, \cite{Tsokaros:2021tsu} and references therein).  

Recently, we have developed a new numerical code for computing equilibriums of 
compact stars associated with mixed poloidal and toroidal magnetic fields 
as a part of our \cocal\ code (Compact Object CALculator) 
\cite{Uryu:2014tda}, \cite{Uryu:2019ckz} (hereafter, Paper I), and \cite{Uryu:2023lgp} (Paper II).  
In our code, exact formulations for the relativistic gravity, the electromagnetic fields, 
and the hydrostationary equilibriums of ideal MHD fluid are incorporated, 
and are consistently solved numerically.  
In \cite{Uryu:2014tda}, and in Papers I and II, we have presented several models of relativistic 
rotating magnetized stars.  Solutions are classified by whether the star is uniformly or 
differentially rotating, and by whether the outside of the star is electromagnetic 
vacuum or the force-free magnetosphere.  Although we have shown only a couple of 
solutions for each model, some of the solutions were actually supramassive.  

In this paper, extending the computations in Papers I and II, we study the supramassive 
as well as hypermassive models whose excess of the mass from the maximum mass of TOV 
solutions are due only to the mixed poloidal and toroidal magnetic fields.  
Among a few models of such magnetic field configurations, 
we consider the cases with the compact stars to be associated with an electromagnetic 
vacuum outside (EMV model), and with a force-free magnetosphere outside (FF model).  
For the former EMV models, the compact star may or may not contain a force-free toroidal 
tunnel inside the star, and its toroidal components should be confined in the 
stellar interior where the ideal magnetohydrodynamic (ideal MHD) condition is satisfied.  
For the latter FF model, the toroidal components of the magnetic fields can extend 
across the interior ideal MHD region to the exterior force-free region (see, Paper II).  
Since our code computes rapidly rotating magnetized stars, non-rotating magnetized 
solutions are obtained practically by minimizing the angular velocity parameter.  
Under the above setups, non-rotating magnetized solutions from lower to higher mass and 
from weak to extremely strong magnetic fields are systematically calculated.  
To our knowledge, the \cocal\ code is unique in calculating such solutions associated 
with mixed poloidal and toroidal magnetic fields under exact treatments of strong gravity 
and strong electromagnetism in magneto-hydrostationary equilibrium.  

Magnetic fields strong enough to largely deform the stellar 
configuration can be as extremely high as $B\sim 10^{18}\,\mbox{G}$ for a typical 
compact star, as such magnetic fields should be close 
to the virial limit, $B\sim 10^{18}\,\mbox{G}\,(R/10\,\mbox{km})^{-2}(M/\Msol)$ 
\cite{Chandrasekhar:1953zz,Suvorov:2021ymy}.  
This is the  magnetic field strength we will see in the interior of our obtained 
solutions below.  
So far it is reported that the surface magnetic field of the magnetar candidate 
of GRB 100625A may be as high as $\sim 10^{17}\,\mbox{G}$ \cite{Rowlinson:2013ue}.  
It is not only interesting to study such an extremely magnetized star as an 
astrophysical object in the present Universe, but 
such solutions may also be useful as initial data of numerical relativity simulations 
to elucidate the properties of the strong gravitational and electromagnetic 
effects on the evolution of compact stars (without the effect of rotation) 
\cite{MRNSsimulationsUIUC,MRNSsimulations}. 
Especially, several general relativistic magnetohydrodynamic (GRMHD) simulations 
for the merger of magnetized binary neutron stars reproduce massive merger remnants 
with local maximum magnetic fields as high as $10^{16.5}$-$10^{17}\,\mbox{G}$ 
\cite{MBNSmergersimulations}. 

This paper is organized as follows.  In Sec.~\ref{sec:Formulation}, 
we review a formulation and numerical method developed in 
our previous works Papers I and II, mentioning that a 
setting of arbitrary function is modified from the previous works.  
In Sec.~\ref{sec:Results}, we show a sequence of solutions systematically 
changing the strength of electromagnetic fields as well as the central density.  
In Sec.~\ref{sec:Discussion}, we comment on some future prospects.

\section{Formulation and numerical computation}
\label{sec:Formulation}

In Paper I, the formulation and numerical method for computing stationary and 
axisymmetric equilibriums of strongly magnetized relativistic rotating stars 
were presented, and in Paper II, they were extended to the case with a 
force-free magnetosphere/magnetotunnel.  Full details of our formulation 
are found in Papers I, II, and \cite{Gourgoulhon:2011gz}.  Our models and assumptions for 
the extremely magnetized compact stars are summarized as follows.  
Hereafter, for index notation for the tensors, the Greek letters 
$\alpha, \beta,...$ denote 4D objects, the Latin lowercase letters $a,b,...$ 
spatial 3D objects, and the Latin uppercase letter $A, B,...$ the meridional 2D objects.  
We use $G=c=M_\odot=4\pi\epsilon_0=1$ units in this paper unless otherwise specified.

\subsection{Summary for formulations}
\label{sec:form}

The compact star is described by the relativistic ideal magnetohydrodynamic 
(ideal MHD) fluid, and may (or may not) have a force-free magnetotunnel inside.  
Outside of the compact star is assumed to be either an electromagnetic vacuum 
or a force-free magnetosphere, and hence either the ideal MHD equations 
(the MHD-Euler equations and the rest mass conservation equation associated 
with the ideal MHD conditions), or the force-free conditions 
are solved simultaneously with Einstein's and Maxwell's field equations.  
For the field equations, we apply 3+1 decomposition with respect to the 
foliation of spatial hypersurfaces $\Sigma$.  Here, we write the 3+1 form of 
the metric, 
\beq
ds^2\,:=\,-\alpha^2dt^2+\gmabd(dx^a+\beta^a dt)(dx^b+\beta^b dt), 
\eeq
where $\alpha$ and $\beta^a$ are the lapse and the shift, and $\gmabd$ 
is a restriction of the projection tensor $\gamma_\albe :=\gabd+n_\alpha n_\beta$ 
to the spatial hypersurface $\Sigma$.  
The hypersurface normal $n_\alpha$ is defined by 
$n_\alpha :=-\alpha\na_\alpha t$.  We further introduce the conformal 
decomposition of the spatial metric $\gmabd:=\psi^4\tgmabd$, and separate 
a flat metric $\fabd$ as $\tgmabd:=\fabd + \habd$.  

Under the assumptions of stationarity and axisymmetry, the system of 
the ideal MHD equations and the force-free conditions is integrated analytically 
to give a system of first integrals and integrability conditions.  
We also assume, for simplicity, that the flow is homentropic, $s=\mbox{constant}$, and 
circular, $u^\alpha = u^t(t^\alpha +\Omega \phi^\alpha)$.  
Then, the following quantities become arbitrary functions of a potential $A_\phi$,  
\beqn
&& 
A_t(A_\phi),\ \ \ \Omega(A_\phi),\ \ \ \Lambda(A_\phi), 
\nonumber 
\\[-2mm]
&& \label{eq:arbfnc}
\\[-2mm]
&& 
\mbox{and}\ \ 
\begin{cases}\ 
[\sqrt{-g}\Lambda_\phi ](A_\phi) \ \  \mbox{for ideal MHD,}
\\[2mm]\ 
[\sqrt{-g}B](A_\phi) \ \ \  \mbox{for force-free,} 
\end{cases}
\nonumber
\eeqn
Here, $t^\alpha$ and $\phi^\alpha$ are the timelike and axial Killing vectors, respectively, 
$u^\alpha$ is the 4-velocity, and $\Omega$ the angular velocity.  
The $t$ and $\phi$ components of the electromagnetic 1-form $A_\alpha$ are written 
$A_t =A_\alpha t^\alpha$ and $A_\phi =A_\alpha \phi^\alpha$.  
The functions $\Lambda(A_\phi)$, and $[\sqrt{-g}\Lambda_\phi](A_\phi)$, are 
the integrability conditions for $t$ and $\phi$ components of MHD-Euler equations, and 
$\epsilon^{AB}B:=F^{AB}$ is the meridional component of the Faraday tensor, 
$\Fabu$.  As discussed in Paper II, the current in the ideal MHD region continues  
smoothly to the force-free region by choosing a common arbitrary function as 
\beq
\frac1{4\pi}[\sqrt{-g}B]'(A_\phi)\,=\, [\sqrt{-g}\Lambda_\phi]'(A_\phi),
\label{eq:idealMHDtoFF}
\eeq
where the prime denotes a derivative with respect to $A_\phi$.  

Finally, choices for the EOS and the rotation laws are also a part of integrability conditions.  
For the EOS, we choose the relativistic polytropic EOS for simplicity, 
\beq
p\,=\,K\rho^\Gamma. 
\label{eq:EOS}
\eeq
For the rotation laws, we set the angular velocity $\Omega(A_\phi)$ to be constant, 
\beq
\Omega(A_\phi)=\Omegac =\mbox{constant}, 
\label{eq:UR}
\eeq
that is, the uniform rotation for the EMV model, and to be 
\beq
\Omega(A_\phi)=\Omegac \Xi'(A_\phi), 
\label{eq:DR}
\eeq
a differential rotation for the FF model, where the function $\Xi'(A_\phi)$ 
will be introduced in the next subsection.  

Obviously, to study the effect of magnetic fields on the mass of compact stars, 
it is desirable to calculate solutions with $\Omegac =0$ (strictly non-rotating solutions).  
To obtain such solutions, however, a certain modification of our code is necessary.  
Instead of modifying the code, we use the same magnetized rotating star code developed in 
our previous works (Papers I and II), and set the parameters of solutions to have $\Omegac$ 
as small as possible, $\Omegac \approx 0$.  
We will show later that the solutions obtained in this way approximate well 
the non-rotating models, and hence regard them as practically non-rotating solutions.

\subsection{Assumptions for arbitrary functions}
\label{sec:MHDfnc}

For the above arbitrary functions (\ref{eq:arbfnc}), we choose the same form 
as in Papers I and II.  For EMV models, 
\beqn
\Lambda &=& - \Lambda_0 \Xi(A_\phi) - \Lambda_1 A_\phi - {\cal E},
\label{eq:MHDfnc_Lambda}
\\[1mm]
A_t &=& - \Omegac A_\phi + C_e, 
\label{eq:MHDfnc_At}
\\[1mm]
&& \!\!\!\!\!\!\!\!\!\!\!\!\!\!\!\!
\sqrt{-g}\Lambda_\phi \,=\, \Lambda_{\phi 0}\, \Xi(A_\phi),
\label{eq:MHDfnc_Lambda_phi}
\\[1mm]
&& \!\!\!\!\!\!\!\!\!\!\!\!\!\!\!\!
\sqrt{-g}B \,=\, 4\pi \Lambda_{\phi 0}\, \Xi(A_\phi),
\label{eq:MHDfnc_B}
\eeqn
where $\Lambda_0$, $\Lambda_1$, ${\cal E}$, $C_e$, and 
$\Lambda_{\phi 0}$ are constant.  For the FF model, 
Eq.~(\ref{eq:MHDfnc_At}) is replaced by 
\beq
A_t \,=\, - \Omegac \Xi(A_\phi) + C_e. 
\label{eq:MHDfnc_At_DR}
\eeq

Parameters 
$\Lambda_0$, $\Lambda_1$, and $\Lambda_{\phi 0}$ are 
prescribed systematically in order to control the strength of electromagnetic 
fields from weak, where the solution agrees well with a non-magnetized 
model, to extremely strong, where the solutions largely deform.  
The other constants ${\cal E}$, $\Omegac$, and $C_e$ are calculated 
from conditions to specify the mass, total angular momentum, and 
charge of a solution.  

Even under the choices of the above relations (\ref{eq:MHDfnc_Lambda})-(\ref{eq:MHDfnc_At_DR}) 
for arbitrary functions (\ref{eq:arbfnc}), 
a form of $\Xi(A_\phi)$ in each relation can still be chosen arbitrarily (hence differently) 
under the restriction (\ref{eq:idealMHDtoFF}).  
In Papers I and II, we have used a so-called sigmoid function for 
the derivative $\Xi'(A_\phi)$ for all relations for simplicity.  
In this work, we modify the function $\Xi'(A_\phi)$ to the fifth-degree polynomial 
which behaves similarly to the sigmoid function as follows, 
\beq
\Xi'(x;b,c)\,:=\,6 \hat{x}^5-5\hat{x}^3+\frac{15}{8}\hat{x} + \frac12, 
\ \ \
\hat{x}:=\frac{x-c}{b}, 
\label{eq:zero1_5th_poly}
\eeq
where $b$ and $c$ are parameters satisfying $0 < b < 1$ and $0 < c < 1$ for 
the EMV model, and $b > 0$ and $c > 0$ for the FF model.  
We set this polynomial $\Xi'(x;b,c)$ to monotonically increase from 0 to 1 in 
an interval $\displaystyle x\in \left[c-\frac{b}{2},c+\frac{b}{2}\right]$.
\!\!\!\footnote{
Note that the derivative of $\Xi'(x;b,c)$ becomes \vfill
$\displaystyle \pa_x \Xi'(x;b,c)=\frac{30}{b^5}\left(x-c-\frac{b}2\right)^2
\left(x-c+\frac{b}2\right)^2$.}
Using this, we set $\Xi'[A_\phi]$, 
\beqn
&&
\Xi'(A_\phi)\,:=\,
\begin{dcases} 
0 & \mbox{for}\qquad\quad\,  \xi \leq -\frac{b}{2},
\\
\Xi'(\xi;b,c)
  & \mbox{for}\ \ -\frac{b}{2} \leq \xi \leq \frac{b}{2},
\ \ \\
1 & \mbox{for}\qquad \frac{b}{2} \leq \xi,
\end{dcases}
\label{eq:zero1_5th}
\\[2mm]
&&\mbox{where} \ \ 
\xi\,:=\,\frac{A_\phi-A_{\phi}^{0}}{A_\phi^{1} - A_{\phi}^{0}}\,-\,c.
\label{eq:zero1_5th_Aphi}
\eeqn
The function $\Xi'(A_\phi)$ varies from 0 to 1 in an interval 
$A_\phi^0 \le A_\phi \le A_\phi^{1}$.  We always set the value of 
$A_\phi$ at the rotation axis ($z$-axis) to be zero.
We also set the constant of integration, 
\beq
\Xi(A_\phi)\,=\,\int\Xi'(A_\phi)dA_\phi\,+\,\mbox{constant}, 
\eeq
to be $\Xi(0)\,=\,0$.  

In previous Papers I and II, we used a sigmoid function for the arbitrary 
function $\Xi'(A_\phi)$.  The new $\Xi'(A_\phi)$, Eqs.~(\ref{eq:zero1_5th}) and 
(\ref{eq:zero1_5th_Aphi}) defined by the polynomial 
(\ref{eq:zero1_5th_poly}), results in practically the same solutions.  
We modify $\Xi'(A_\phi)$ because the sigmoid function does not strictly 
become 0 or 1 in its asymptotic region.

\subsection{Model parameters}
\label{sec:model_para}

A solution of a magnetized relativistic star is specified with a set of parameters, 
$\left\{\Lambda_0, \Lambda_1, \Lambda_{\phi 0}, {\cal E}, \Omegac, C_e \right\}$, 
appearing in Eqs.~(\ref{eq:MHDfnc_Lambda})-(\ref{eq:MHDfnc_At_DR}), and a set of parameters 
$\{A_{\phi}^{0}, A_\phi^{1}, b, c \}$ in the function $\Xi(A_\phi)$ as 
Eqs.~(\ref{eq:zero1_5th_poly})-(\ref{eq:zero1_5th_Aphi}).  
The latter set of parameters can be prescribed differently in each $\Xi(A_\phi)$ appearing in 
Eqs.~(\ref{eq:MHDfnc_Lambda})-(\ref{eq:MHDfnc_At_DR}) (under a restriction (\ref{eq:idealMHDtoFF})), 
but they are chosen to be the same except for the $\Xi(A_\phi)$ in differential rotation law 
(\ref{eq:DR}) (and hence (\ref{eq:MHDfnc_At_DR})).  

For the case of the EMV model, the latter set $\{A_{\phi}^{0}, A_\phi^{1}, b, c \}$ 
are chosen as in Papers I and II, namely, 
$(A_\phi^{0}, A_\phi^{1})=(A_{\phi, \rm S}^{\rm max},A_\phi^{\rm max})$ 
and $(b,c)=$ $(0.2,0.5)$, 
where $A_{\phi, \rm S}^{\rm max}$ is the maximum value of $A_\phi$ on the stellar surface, 
and $A_\phi^{\rm max}$ the maximum value of $A_\phi$ on the stellar support.  
With the choice $A_\phi^0 = A_{\phi, \rm S}^{\rm max}$, 
the toroidal component of the magnetic fields is confined inside the star.
\!\!\!\footnote{In Paper II, we referred to this type of solution with the electromagnetic 
vacuum outside, with the force-free magnetotunnel, and with the uniform rotation, 
EV-MT-UR type solutions.}

For the case of the FF model, the latter set $\{A_{\phi}^{0}, A_\phi^{1}, b, c \}$
for Eqs.(\ref{eq:MHDfnc_Lambda}), (\ref{eq:MHDfnc_Lambda_phi}), and (\ref{eq:MHDfnc_B}) are chosen as 
$(A_\phi^{0}, A_\phi^{1})=(0.3A_{\phi, \rm S}^{\rm max},1.7A_\phi^{\rm max})$ and 
$(b,c)=$ $(1.2,0.8)$, and for the rotation law (\ref{eq:DR}) and (\ref{eq:MHDfnc_At_DR}), they are  
$(A_\phi^{0}, A_\phi^{1})=(0.0,A_\phi^{\rm max})$ and $(b,c)=$ $(0.1,0.15)$.

Intensities of the electromagnetic fields inside and outside of 
the compact stars are controlled by the first three parameters of 
the former set, 
$\{\Lambda_0, \Lambda_1, \Lambda_{\phi 0}, {\cal E}, \Omegac, C_e \}$.  
As we have mentioned, we consider non-rotating solutions $\Omegac \approx. 0$.  
The parameter $C_e$ is evaluated from a condition to vanish the asymptotic 
charge $Q$.  The parameter ${\cal E}$ is evaluated from 
a condition to set a given value for a ratio of the central pressure and density 
$\prhoc$, which determines the mass of the compact star.  
Therefore, free parameters in actual computations are 
$\{\Lambda_0, \Lambda_1, \Lambda_{\phi 0}, \prhoc\}$.  

For the polytropic EOS (\ref{eq:EOS}), we choose the polytropic 
index $\Gamma=2$, and the polytropic constant $K$ is chosen such 
that the value of $M_0$ becomes $M_0=1.5$ at the compactness 
$M/R=0.2$ for a spherically symmetric solution in $G=c=M_\odot=1$ units.  
Here, $M_0$ is the rest mass, $M$ the gravitational mass, and $R$ the 
circumferential radius.  The maximum-mass model and conversion 
factors to cgs units are shown in Table \ref{tab:TOV_solutions}. 

\begin{table}
\caption{
Quantities at the maximum-mass model of Tolman-Oppenheimer-Volkoff (TOV) 
solutions with the polytropic EOS (\ref{eq:EOS}) in $G=c=M_\odot=1$ units.  
$\presc$ and $\rhoc$ are the pressure and the rest-mass density at the center, 
$M_0$ is the rest mass, $M$ the gravitational mass, and 
$M/R$ the compactness (a ratio of the gravitational mass to 
the circumferential radius).  
To convert a unit of $\rhoc$ to cgs, multiply the values by 
$\Msol(G\Msol/c^2)^{-3} \approx. 6.172716\times 10^{17} {\rm g}\ {\rm cm}^{-3}$.}  
\label{tab:TOV_solutions}
\begin{tabular}{cccccc}
\hline
$\Gamma$ & $\prhoc$ & $\rhoc$ & $M_0$ & $M$ & $M/R$ \\
\hline
$2$ & $0.318244$ & $0.00448412$ & $1.51524$ & $1.37931$ & $0.214440$\\
\hline
\end{tabular}
\end{table}

\subsection{Summary for numerical computation}
\label{sec:num_comp}

Our numerical method for computing magnetized relativistic stars is 
detailed in Papers I and II, as well as in our previous papers 
on a development of \cocal\ code \cite{cocal}.  A system of equations includes 
elliptic partial differential equations for the gravitational 
and the electromagnetic field variables, and algebraic relations 
for the fluid variables.  Those equations are discretized on 
spherical coordinates, and solved iteratively, writing the elliptic 
equations in the integral form using the Green's function.  
The multipole expansion is applied to evaluate the Green's function, 
where the multipoles whose order is higher than $L=60$ are truncated.  
The resolution of finite difference is set accordingly to resolve 
the highest multipoles, which is the same as the highest resolution 
used in Papers I and II.  
The numbers of grid points and other grid parameters, 
also used in the present computations, 
are reproduced in Table \ref{tab:RNS_grids} marked as SE3tp.  

For a typical run on a workstation, it uses about 40 GB memory, and 
takes 8.3 min, 12.5 min, and 16 min, per 1 iteration using a single core of 
Intel Xeon w5-3525 CPU(3.20GHz), 
Intel Xeon Platinum 8380 CPU(2.30GHz), and 
Intel Xeon Gold 6254 CPU(3.10GHz), respectively, 
and it takes about 600 iterations for a convergence.  
These systems are fast enough to compute an equilibrium solution or 
an initial dataset on a single time slice, however, it is desirable to 
speedup the code for systematic computations in a wider parameter range. 
Recently, parallelized versions of \cocal\ codes have been developed for   
(non-magnetized) rotating compact star codes or binary neutron star codes 
\cite{Boukas:2023ckb}, which reached speedups of $\sim 20$ times faster 
than the serial code using $\sim 40$ cores ($\sim 10$ times faster 
using $\sim 20$ cores) on the same kind of workstations as the above.  
A parallelization of the present magnetized rotating star code is in progress, 
which is expected to achieve a similar speedup.  
\begin{table}
\caption{Grid parameters used for computing magnetized 
rotating compact stars.  
Normalized radial coordinates $r_a$, $r_b$, and $r_c$ are in the 
unit of equatorial radius $R_0$ in the coordinate length.
}
\label{tab:RNS_grids}
\begin{tabular}{cccccccccl}
\hline
\qquad Type  & $r_a$ & $r_b$ & $r_c$ & $\Nrf$ & $\Nrm$ & $N_r$ & $N_\theta$ & $N_\phi$ & $L$ \\
\qquad SE3tp & 0.0 & $10^6$ & 1.1  & 160 & 176  & 384  & 384  & 72 & 60 \\
\hline
\multicolumn{10}{l}{\,$r_a$ : Radial coordinate where the radial grids starts.} \\
\multicolumn{10}{l}{\,$r_b$ : Radial coordinate where the radial grids ends.} \\
\multicolumn{10}{l}{\,$r_c$ : Radial coordinate between $r_a$ and $r_b$ where} \\
\multicolumn{10}{l}{\qquad the radial grid spacing changes.} \\
\multicolumn{10}{l}{$N_r$\ \ : Number of intervals $\Dl r_i$ in $r \in[r_a,r_b]$.} \\
\multicolumn{10}{l}{$\Nrf$\ \ : Number of intervals $\Dl r_i$ in $r \in[r_a,1]$.} \\
\multicolumn{10}{l}{$\Nrm$ : Number of intervals $\Dl r_i$ in $r \in[r_a,r_{c}]$.} \\
\multicolumn{10}{l}{$N_{\theta}$\ \ : Number of intervals $\Dl \theta_j$ in $\theta\in[0,\pi]$.} \\
\multicolumn{10}{l}{$N_{\phi}$\ \ : Number of intervals $\Dl \phi_k$ in $\phi\in[0,2\pi]$.} \\
\multicolumn{10}{l}{$L$\ \ \ \ : Order of included multipoles.} \\
\hline
\end{tabular}  
\end{table}
\begin{table}
\caption{Parameters of arbitrary functions in the integrability 
conditions (\ref{eq:MHDfnc_Lambda})--(\ref{eq:MHDfnc_B}).  
}
\label{tab:MHDfnc_param}
\begin{tabular}{lccc}
\hline
Models & $\Lambda_0$ & $\Lambda_1\ \ $ & $\Lambda_{\phi0}$ \\
\hline
EMV$^-$I   & $-2.4\quad$ & $0.24\quad$ & $1.5$  \\
EMV$^-$II  & $-3.2\quad$ & $0.32\quad$ & $2.0$  \\
EMV$^-$III & $-4.0\quad$ & $0.40\quad$ & $2.5$ \\
EMV$^-$IV  & $-4.8\quad$ & $0.48\quad$ & $3.0$  \\
\hline
EMV$^+$  & $\quad 0.4\quad$ & $0.30$--$0.567\quad$ & $1.0$  \\
\hline
FF  & $\quad 4.8\quad$ & $0.30\quad$ & $1.2$  \\
\hline
\end{tabular}
\end{table}

\section{Results}
\label{sec:Results}

\subsection{Parameters and caveats for the non-rotating magnetized solutions}
\label{sec:non-rotatingMRNS}

Because of the limitation of computing resources, we have investigated 
only a part of the whole parameter space of the four parameters, 
$\{\Lambda_0, \Lambda_1, \Lambda_{\phi 0}, \prhoc\}$.  
From our numerical experiments, we found that some combinations of values for 
magnetic field parameters, $\{\Lambda_0, \Lambda_1, \Lambda_{\phi 0}\}$, result in 
a strong toroidal component comparable to a poloidal one, that is, they reproduce 
solutions associated with mixed poloidal and toroidal fields, so-called 
``twisted-torus solutions''.  
We present three classes of such solutions, two of which belong to EMV models, 
and one to FF models.  We denote the EMV models as EMV$^+$ and EMV$^-$, 
corresponding to a positive or negative value of the parameter $\Lambda_0$, 
respectively.  For the FF model, we will show a model with $\Lambda_0>0$.  

For the EMV$^-$ models, we systematically change the 
parameters as tabulated in Table \ref{tab:MHDfnc_param}.  
The strengths of the magnetic fields from weaker to stronger correspond 
to EMV$^-$I to IV in Table \ref{tab:MHDfnc_param}.  
For each EMV$^-$ model, we set the ratio $\prhoc$ to the value of 
the TOV solution to be $M_0=1.25, 1.3, 1.35, 1.4, 1.45, 1.5$ and then 
$\prhoc=0.3, 0.35, 0.4$, where $\prhoc=0.3$ is the closest to 
that of the maximum mass model of the TOV solution (see Table \ref{tab:TOV_solutions}).  

As our code is to compute \emph{rotating} magnetized relativistic stars, 
the ratio of the polar to the equatorial radius is 
used for a parameter to fix the angular velocity $\Omegac$.  
For a convergence of a solution, the polar radius $R_z$ and 
the equatorial radius $R_0$ are placed on 
certain grid points of $z$ and $x$ coordinates, respectively, and so the ratio 
can be varied only every $1/\Nrf$, where $\Nrf +1$ is the number of grid points 
along the equatorial radius (from the stellar center to the surface).  
Therefore, $\Omegac$ cannot be strictly set $\Omegac=0$, but it is set as small as 
possible in the following computations of (almost) non-rotating solutions.  

As we will see below, when the magnetic fields are weak enough, as with EMV$^-$I models, 
solutions with the slowest rotation with $R_z/R_0=(\Nrf-1)/\Nrf$ are obtained 
which result in our resolution $\ToverW <10^{-3}$, where $\ToverW$ is the ratio of 
the relativistic kinetic to gravitational terms appearing in the relativistic 
virial relation \cite{virial}, 
\beqn
&&-\frac1{8\pi}\int_\Sigma \gamma^\albe\left(\Gabd-8\pi\Tabd\right)dV
\nonumber\\
&&=\,2{\cal T}+3\Pi+{\cal M}+{\cal W}+\Madm - \MK \,=\,0,  \quad
\label{eq:virial_relation}
\eeqn
where $\cal T$, $\Pi$, ${\cal M}$, and ${\cal W}$, are kinetic, 
pressure (internal energy), electromagnetic, and gravitational terms, respectively, 
whose concrete forms are found in Paper I.  
As a result of an equality of ADM and Komar masses $\Madm=\MK$ \cite{virial} 
for an asymptotically flat spacetime, we have 
\beq 
I_{\rm vir}\,:=\,\left|2{\cal T}+3\Pi+{\cal M}+{\cal W}\right|\,=\,0.
\label{eq:virial}
\eeq
For the EMV$^-$II-IV models, the equilibrium configurations are substantially deformed.  
We set $R_z/R_0$ as large as possible and confirm that 
$\ToverW <2\times 10^{-3}$ is always satisfied except for a few solutions 
of the FF model which are $\ToverW <9\times 10^{-3}$.  
In Appendix \ref{sec:App_physq}, we present the physical quantities of selected solutions, including 
the negligible rotation parameters such as the angular velocity and the angular momentum.  
Among the constructed non-rotating solutions in this parameter range of EMV$^-$ models, 
we found supramassive solutions but not hypermassive solutions.  

For the EMV$^+$ models, we set the parameters to compute from weak to strong magnetic field 
solutions similarly to EMV$^-$ models, but change the parameter $\Lambda_1$ only 
in the range shown in Table \ref{tab:MHDfnc_param}.  We set the ratio $\prhoc$ to 
the value of the TOV solution to be $M_0=1.25, 1.3, 1.35, 1.4, 1.45, 1.5$ and then $\prhoc=0.3$, 
and the parameter $\Lambda_1$ is chosen appropriately depending on the value of $\prhoc$ 
as explained in detail below.  Otherwise, our computations proceed the same as with EMV$^-$ models.  
For a fixed $\prhoc$, with increasing $\Lambda_1$, the mass of the magnetized star 
increases substantially so that hypermassive solutions are obtained.  

Finally for the FF model, we choose only one parameter set, shown in 
Table \ref{tab:MHDfnc_param}, and we set the ratio $\prhoc$ to the value of 
the TOV solution to be $M_0=1.25, 1.3, 1.35, 1.4, 1.45, 1.5$ and then 
$\prhoc=0.3, 0.35, 0.4$.  For this model, magnetically hypermassive solutions are 
systematically obtained.  

In the following, we show as an example the structure of a massive solution of such 
magnetized equilibrium for each EMV$^-$, EMV$^+$ and FF model, then present 
the solution sequences of magnetically supramassive and hypermassive compact stars.

\begin{figure*}
\begin{center}
\includegraphics[height=45mm]{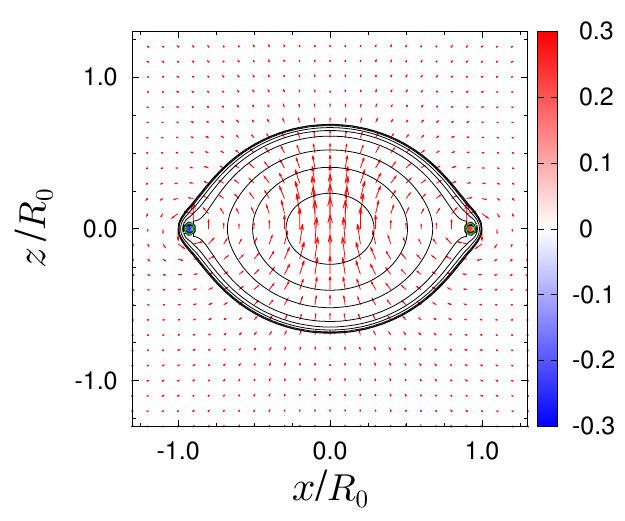}\hspace{5mm}
\includegraphics[height=45mm]{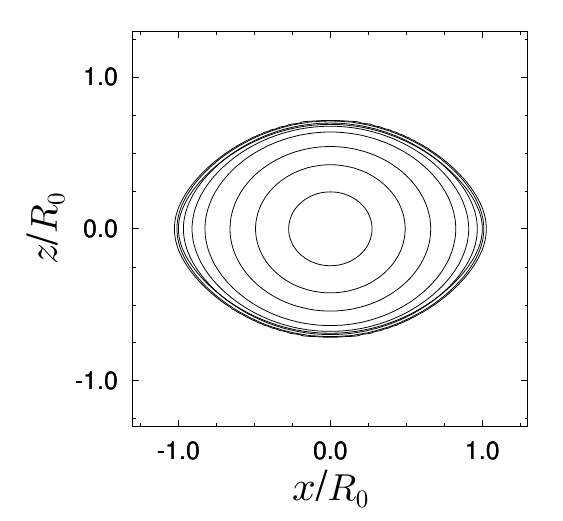}\hspace{5mm}
\includegraphics[height=45mm]{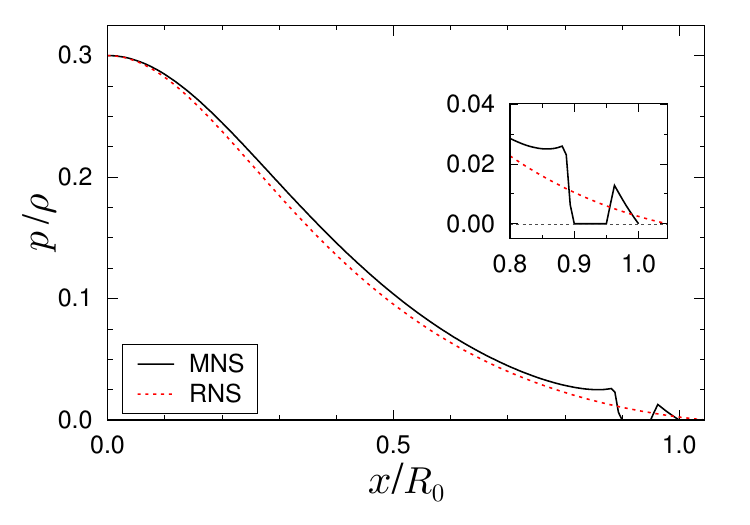}
\\
\includegraphics[height=45mm]{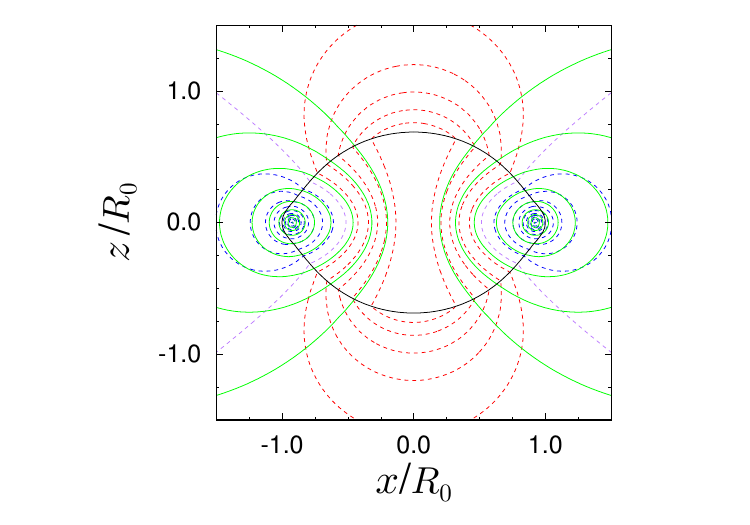}\hspace{5mm}
\includegraphics[height=45mm]{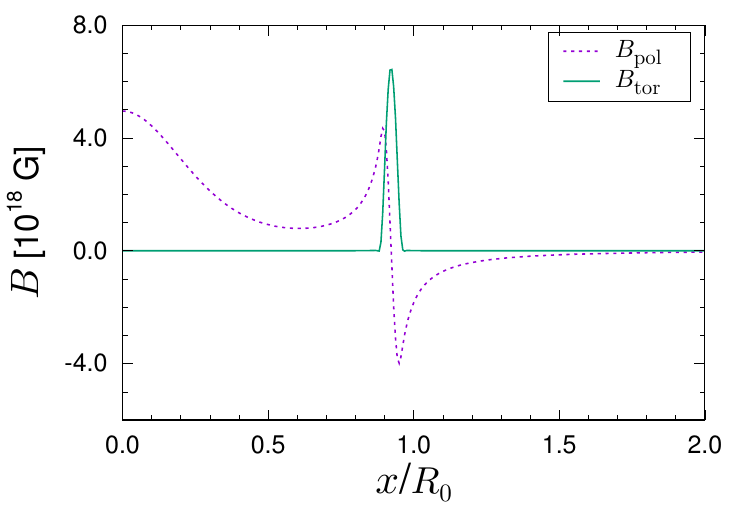}
\\
\includegraphics[height=37mm]{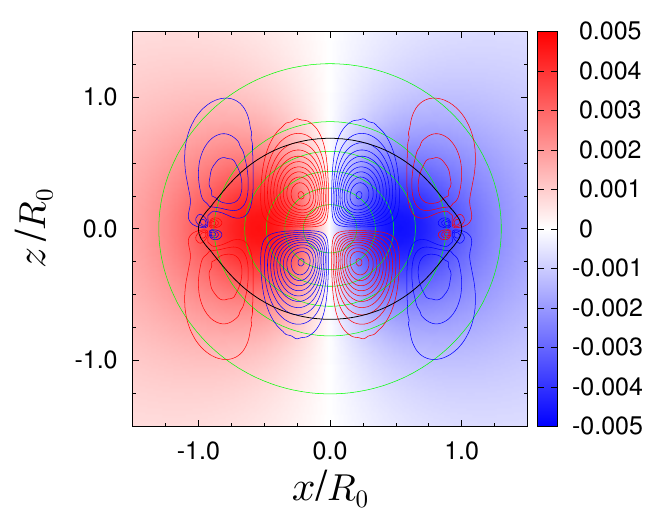}
\includegraphics[height=37mm]{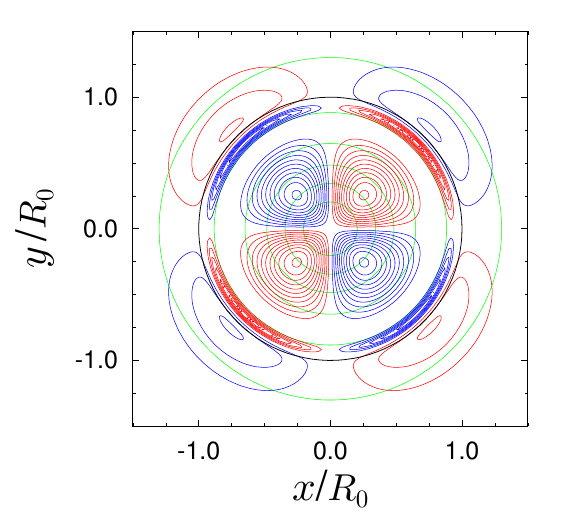}
\includegraphics[height=37mm]{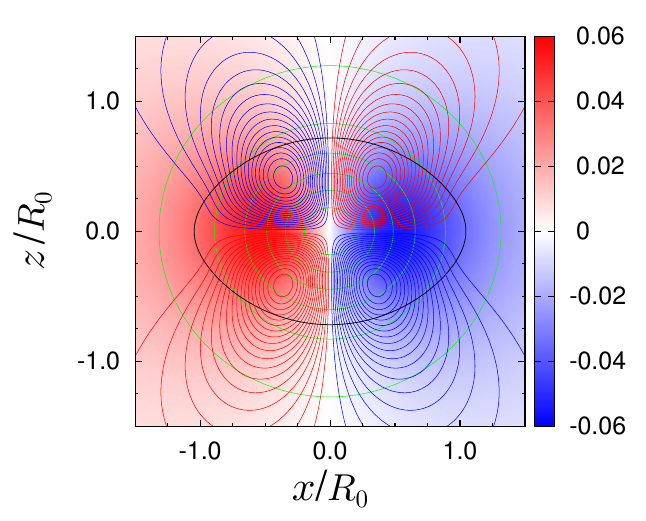}
\includegraphics[height=37mm]{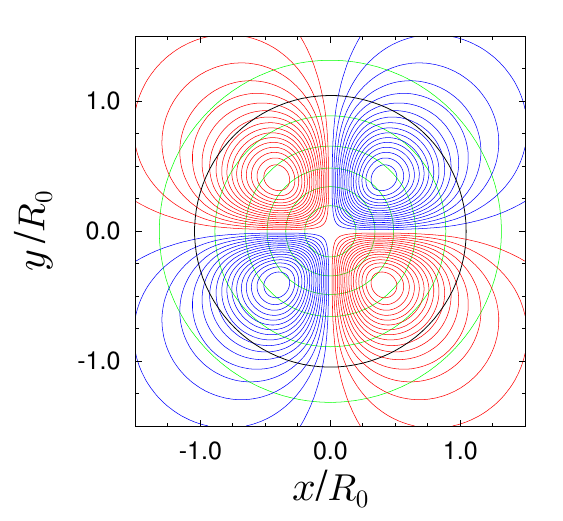}
\caption{
A (non-rotating) extremely magnetized supramassive compact star associated 
with mixed poloidal and toroidal magnetic fields (EMV$^-$IV model) and a uniformly rotating (non-magnetized) 
compact star are compared for the models with the same axis ratio $R_z/R_0=0.6875$ and 
the central $\prhoc=0.3$.  
First row, left panel: contours of $p/\rho$ (black closed curves), 
vector plots of poloidal magnetic field (red arrows), color density map 
for the toroidal magnetic fields (red and blue), and the boundary of 
the magnetotunnel (green circles) for the magnetized star are shown.  
Middle panel: contours of $p/\rho$ for the rotating star are shown.  
The contours are drawn at $p/\rho=0.001,0.002,0.005,0.01,0.02,0.05,0.1,0.2$.  
Right panel: the profiles of $p/\rho$ for both stars are plotted along 
the normalized equatorial coordinate $x/R_0$, and a close-up of the region 
$x/R_0 \geq 0.8$ is shown in an inset.  
Second row, 
left panel: the contours of the components of electromagnetic 1-form
$A_\phi$ (green curves), the contours of $A_t$ (dashed red (positive) 
purple (zero), blue (negative)) are shown for the magnetized star.  
The black closed curve is the surface of the star.  
Right panel: components of the magnetic fields, $B_{\rm pol} = F_{xy}$ 
(dashed purple curve) and $B_{\rm tor} = -F_{xz}$ (dark green curve) 
are plotted with respect to the normalized equatorial radius $x/R_0$.  
Third row, 
left two panels: the metric potentials for the magnetized star 
in $xz$ and $xy$ planes are shown in the left and right panels, respectively, 
with the contours of $\psi$ (green closed curves), 
the color density map for $\tbeta_y$ (red and blue, left panel only), 
the contours of $h_{xz}$ (left panel) and $h_{xy}$ (right panel) (red and blue curves), 
and the surface of the star (black closed curve)."
Right two panels: the same as the left two panels but for the uniformly rotating 
(non-magnetized) star.  
The coordinate lengths in all panels are normalized by the equatorial radius $R_0$ 
of the magnetized star, which is about $4.5\%$ smaller than that of the rotating star.  
}  
\label{fig:MNS-RNS}
\end{center}
\end{figure*}
\begin{figure*}
\begin{center}
\includegraphics[height=42mm]{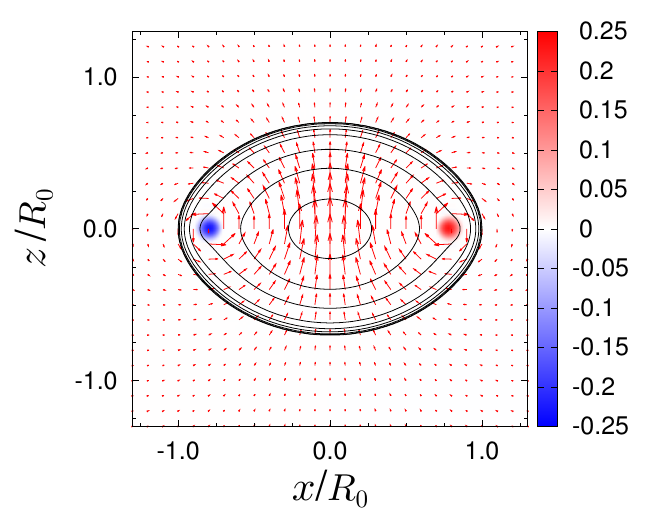}
\includegraphics[height=42mm]{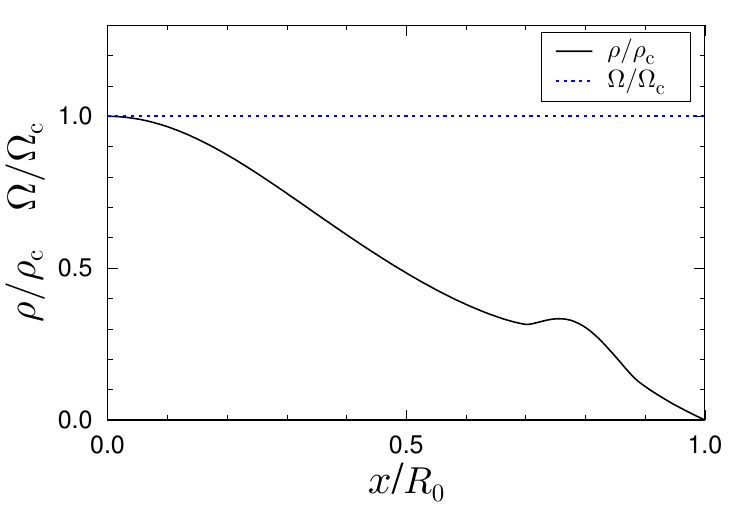}
\includegraphics[height=42mm]{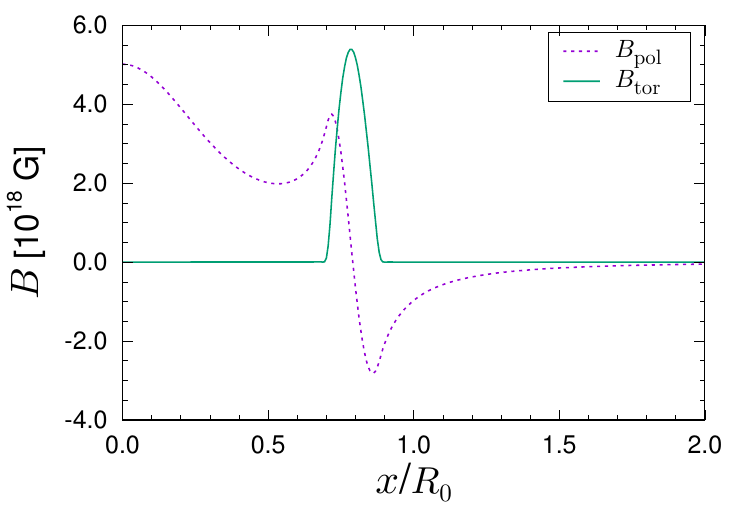}
\\
\includegraphics[height=42mm]{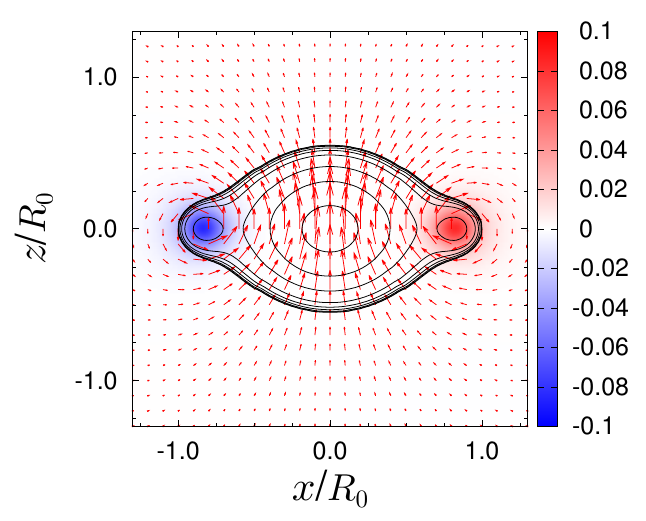}
\includegraphics[height=42mm]{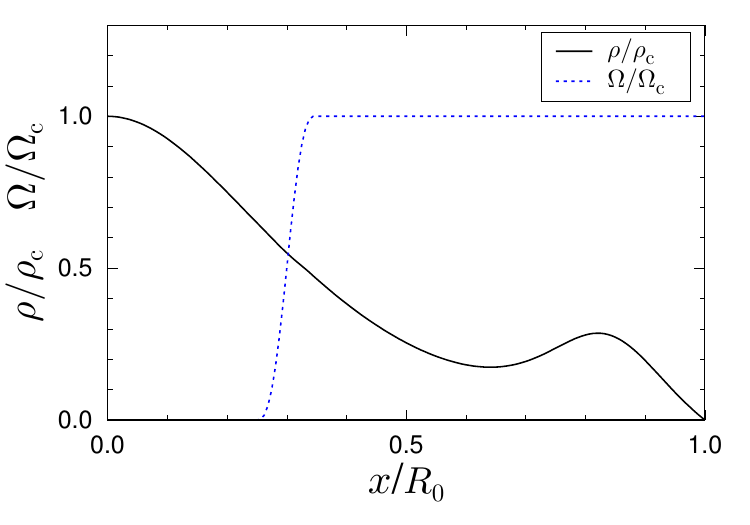}
\includegraphics[height=42mm]{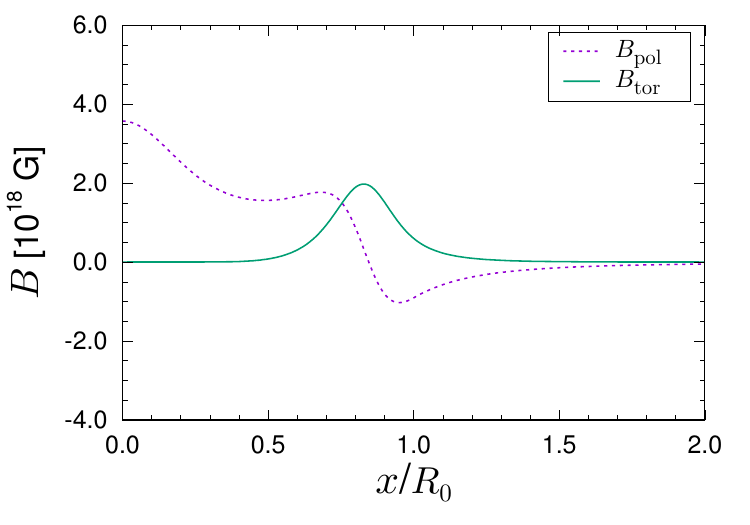}
\\
\includegraphics[height=36mm]{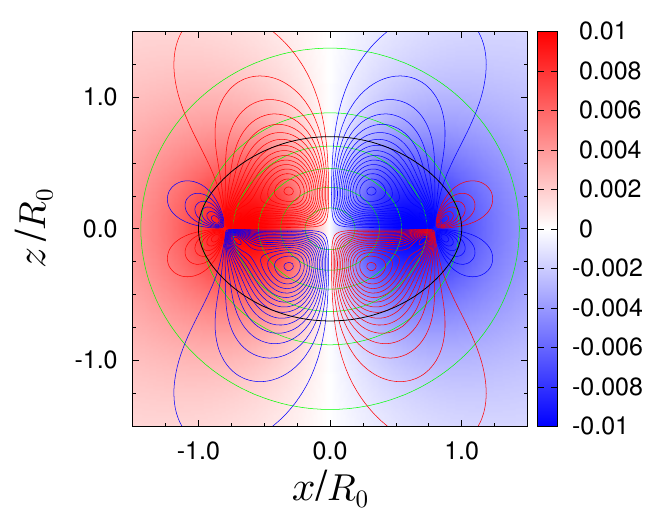}
\includegraphics[height=36mm]{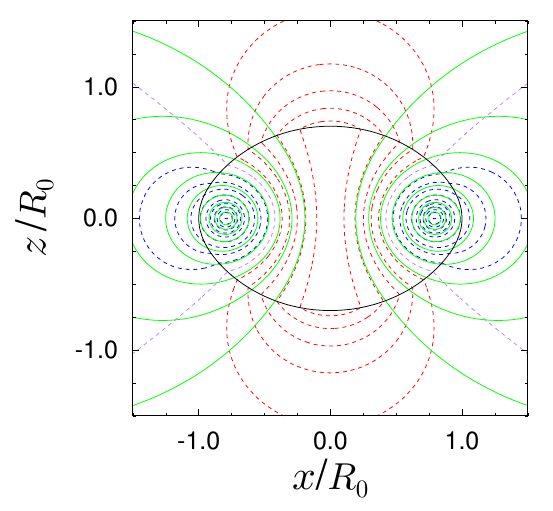}
\includegraphics[height=36mm]{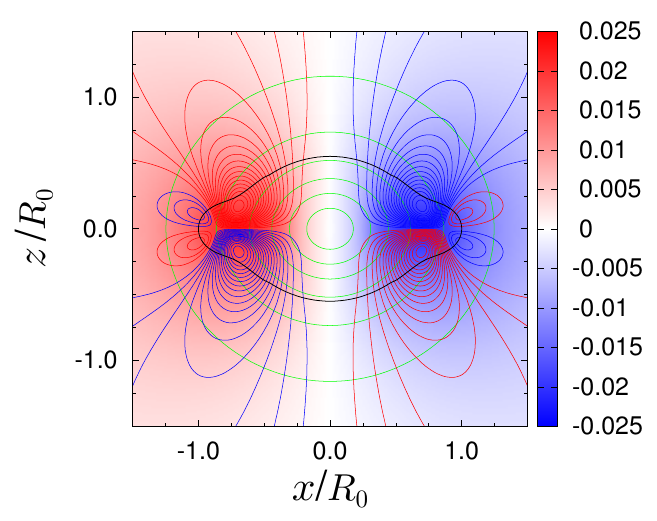}
\includegraphics[height=36mm]{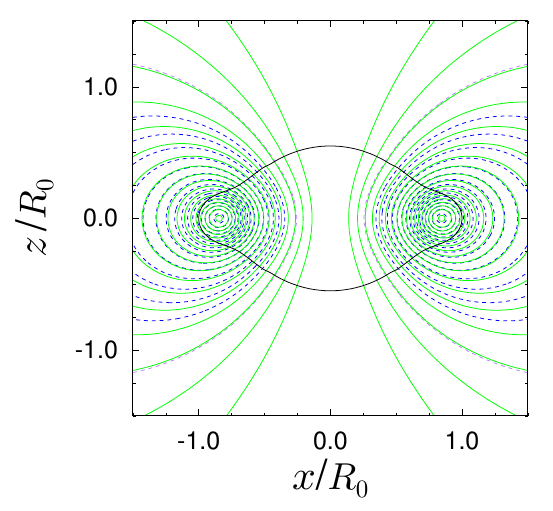}
\\
\caption{
Solutions for magnetically hypermassive compact stars associated 
with an electromagnetic vacuum outside (EMV$^+$ model) and with a force-free 
magnetosphere outside (FF model).  
First row: the EMV$^+$ model with $\Lambda_1=0.432$, $\prhoc=0.2558203$ ($M_0=1.5$), 
and $R_z/R_0=0.7$.  
Left panel: same as the left panel of the first row of Fig.\ref{fig:MNS-RNS}.
Middle panel: the normalized rest-mass density $\rho/\rhoc$ (black curve) and 
the normalized angular velocity $\Omega/\Omegac$ (blue dotted curve) are plotted 
along the equatorial radius ($x$-axis).  
Right panel: same as the right panel of the second row of Fig.~\ref{fig:MNS-RNS}.
Second row: the panels are the same as the first row but for the FF model 
with $\prhoc=0.2558203$ ($M_0=1.5$), and $R_z/R_0=0.55$.  
Third row: the first and second panels from the left are those of the EMV$^+$ model 
of the first row, and the third and fourth the FF model of the second row.  
First panel from left: same as the first (left) panel of the third row of 
Fig.~\ref{fig:MNS-RNS}.
Second panel: same as the left panel of the second row of Fig.~\ref{fig:MNS-RNS}.
Third and fourth panels: same as the first and the second panels, respectively.
}  
\label{fig:EMV+FF}
\end{center}
\end{figure*}

\subsection{Solutions of magnetically supramassive and hypermassive compact stars}

\subsubsection{Structural comparison of purely magnetized and purely rotating compact stars}
\label{sec:comparison}

In Fig.~\ref{fig:MNS-RNS}, we show configurations for (non-rotating) magnetically supramassive  
compact star, model EMV$^-$IV, and a uniformly rotating (non-magnetized) supramassive compact star, 
setting the deformation parameter (the axis ratio in a coordinate length) 
$R_z/R_0 = 0.6875$, and the ratio $\prhoc=0.3$ to be the same for comparison.  
The uniformly rotating solution is calculated using the \cocal\ code solving 
the same waveless formulation under the same coordinate conditions 
(the maximal slicing and the Dirac gauge conditions) as the magnetized solution.  

In all panels of Fig.~\ref{fig:MNS-RNS}, coordinate lengths $(x, y, z)$ are all normalized by 
the equatorial radius $R_0$ of the magnetized solution.  It is noticed that the size of 
the non-magnetized rotating solution is about $4.5\%$ larger in coordinate than the magnetized model.  
In the first row of the Fig.~\ref{fig:MNS-RNS}, the contours of $p/\rho$ and its profiles along 
the $x$-axis are shown for the magnetized solution as well as for the rotating solution, and 
in the second row, the contours of the components of electromagnetic 1-form $A_t$ and $A_\phi$, 
and the plots of the components of poloidal and toroidal magnetic fields $\Bpol:=F_{xy}$ and 
$\Btor:=-F_{xz}$ are presented.  For the same amount of deformation, the magnetized configuration 
is slightly more compact (that is, the EOS becomes effectively stiffer) than the rotating one, 
except for the magnetotunnel region near the equatorial surface where the magnetic fields expel 
the matter.  

For these extremely strong magnetic fields, a qualitative difference can be seen in the metric 
components.  In the bottom row of Fig.~\ref{fig:MNS-RNS}, contours of the metric components 
$h_{xz}$ in the $xz$-plane, and $h_{xy}$ in the $xy$-plane are drawn.  Because of the strong magnetic fields 
near the equatorial surface, the contours for $h_{xz}$ and $h_{xy}$ for the magnetized model 
(the first and second panels from the left of the third row) have more complex structures 
than those for the rotating model (the third and fourth panels from the left ).  
The density map for the component $\tbeta_y$ is also shown in the first and the third panels 
of the third row ($xz$-plane).  The effect of magnetic fields does not seem to be large enough 
to affect this component.  Note that because of the lack of significant rotation of 
the magnetized model, $\tbeta_y$ is about an order of magnitude smaller than that of the rotating 
model.

\subsubsection{Structures of hypermassive solutions}
\label{sec:hypermassive}

In Fig.~\ref{fig:EMV+FF}, we show solutions of magnetically hypermassive compact stars for 
the EMV$^+$ and FF models.  Model parameters, other than Table \ref{tab:MHDfnc_param}, 
are set to be $\Lambda_1 = 0.432$, $\prhoc=0.2558203$ ($M_0=1.5$), and the axis ratio 
(in coordinate length ) $R_z/R_0=0.7$ for the EMV$^+$ model, 
and $\prhoc=0.2558203$ ($M_0=1.5$), and $R_z/R_0=0.55$ for the FF model.  
As seen in the contours of $p/\rho$ and the density profile $\rho/\rhoc$ along the (normalized) 
$x$-axis, shown in the left and middle panels of the first and second rows, 
there are toroidal concentrations of matter near the equatorial surface for these solutions with 
$\Lambda_0 >0$ due to extremely strong magnetic fields.  This is in contrast with 
the EMV$^-$ model $(\Lambda_0 < 0)$ that has a force-free tunnel.  

A difference between 
the electromagnetic vacuum and the magnetosphere outside of the star can be seen clearly 
in the contours of $A_t$ and $A_\phi$ in the second and fourth panels of the third row.  
As mentioned in Paper II, the toroidal components of magnetic fields ought to be confined inside the 
star (ideal MHD region) when the outside is the electromagnetic vacuum.  For the stars surrounded 
by the force-free magnetosphere, the toroidal component may or may not extend across the surface 
depending on the choice of arbitrary functions and their parameters.  With a choice of parameters 
discussed in Sec.~\ref{sec:model_para}, the region of the toroidal magnetic field is widely extended 
across the surface in our FF model, as shown in the left and right panels of the second row 
in Fig.~\ref{fig:EMV+FF}.  

The masses of EMV$^+$ and FF models in Fig.~\ref{fig:EMV+FF} are $\Madm = 1.6031$ and $1.8132$, respectively.  
The former mass is about 16\% above the maximum mass of TOV solution and slightly above the maximum of 
the uniformly rotating (non-magnetized) star, while the latter is about 31\% heavier than the TOV maximum.
It is noticed in the right panels of the first and second rows in Fig.~\ref{fig:EMV+FF} 
that the maximums of the $\Bpol$ and $\Btor$ of the EMV$^+$ model are much larger than those of 
the FF model.  The FF model with a smaller magnetic field is much heavier because its radius is larger than 
the EMV$^+$ model.  In proper length, it is $\bar{R}_0=17.217$km for FF the model, and 
$\bar{R}_0=13.437$km the EMV$^+$ model.

\begin{figure}
\begin{center}
\includegraphics[height=60mm]{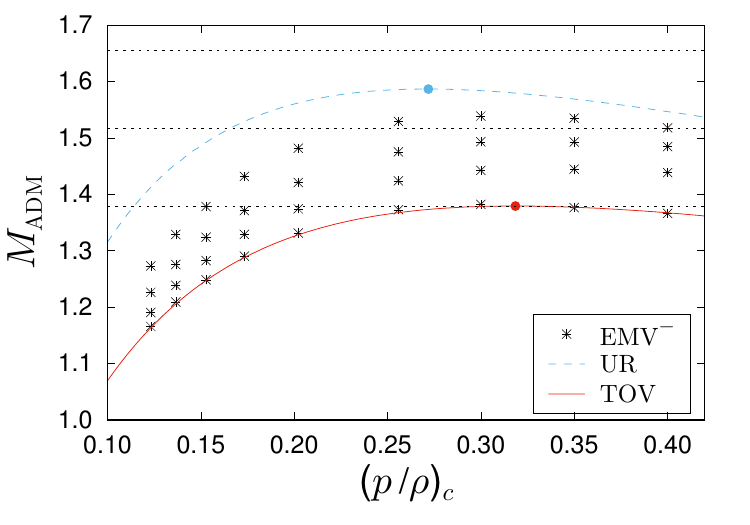}
\includegraphics[height=60mm]{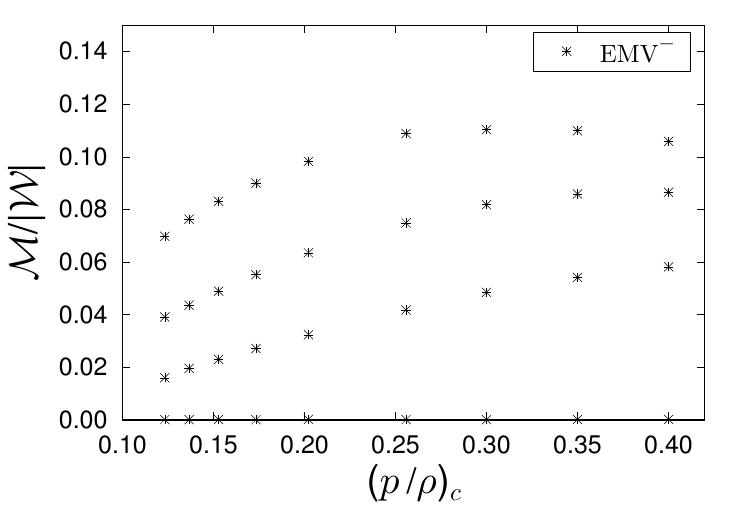}
\\
\caption{Sequence of solutions for EMV$^-$I--IV models.
Top panel: plots of the ADM mass $M$ with respect to the ratio $\prhoc$ 
of magnetized compact stars (plus-crosses, EMV$^-$).  Solid (red, TOV) and dashed (blue, UR) curves 
correspond, respectively, to the solution sequences of the TOV equation and to the maximally 
rotating (Kepler rotation) models of uniformly rotating relativistic stars with the same EOS.  
The maximum mass of each curve is shown with a filled circle, and the horizontal dashed lines 
are drawn at the maximum mass of the TOV solutions and at the 10\% and 20\% higher values from 
bottom to top, respectively.  
Bottom panel: plots of the ratio of magnetic to gravitational energy $\MoverW$ 
with respect to $\prhoc$ for the same EMV$^-$I--IV models.  
In both panels, data points (plus-crosses) from bottom to top at the same $\prhoc$ correspond to 
EMV$^-$I--IV models, respectively.  
}  
\label{fig:M_prhoc}
\end{center}
\end{figure}

\begin{figure}
\begin{center}
\includegraphics[height=60mm]{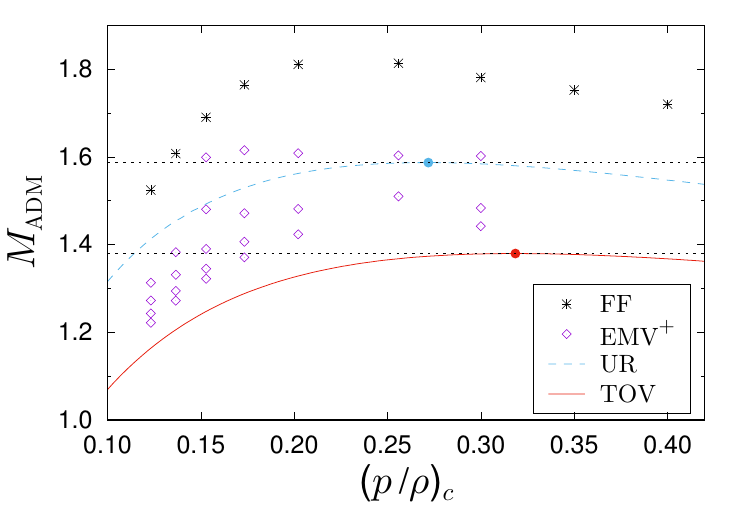}
\includegraphics[height=60mm]{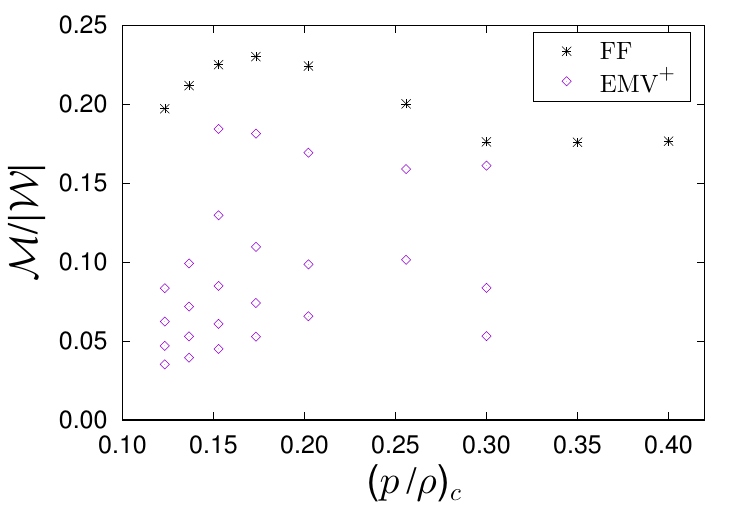}
\\
\caption{The same as Fig.~\ref{fig:M_prhoc}, but for the EMV$^+$ (diamonds purple) and FF (plus-crosses) 
models that include hypermassive solutions.  In the top panel, the horizontal dashed lines are drawn 
at values of the maximum mass of the TOV solutions (red) and at the maximum mass of the uniformly 
rotating solutions (blue).
}  
\label{fig:M_prhoc_hyper}
\end{center}
\end{figure}

\subsection{Magnetically supramassive and hypermassive solutions}
\label{sec:SMHMsequence}

In the top panel of Fig.~\ref{fig:M_prhoc}, we show the $\Madm$--$\prhoc$ relation 
for EMV$^-$I--IV models.  
It is confirmed that the masses of strongly magnetized compact stars can exceed 
the maximum mass of the TOV solutions of the same EOS when the magnetic 
field and the central $\prhoc$ are high enough.  
The excess of mass can be as large as that due to uniform rotation, 
but its maximum value does not surpass that of the maximally rotating (Kepler rotation) 
model in the selected range of parameters.  Therefore, a hypermassive solution 
is not found for the EMV$^-$ models within the searched parameter range.  
Note that EMV$^-$I models agree very well with the TOV solutions, showing that 
those solutions with small $\ToverW$ are regarded well as non-rotating solutions.  

In the bottom panel, we plot the ratio of magnetic to gravitational energy $\MoverW$ 
with respect to $\prhoc$ for the same set of solutions.  Their values 
reach around $\MoverW \sim 0.02-0.1$ which are comparable to typical values of $2\ToverW$ 
of rapidly rotating compact stars (see Eq.~(\ref{eq:virial})).  A value of $\ToverW$ of 
the compared rotating solution in Fig.~\ref{fig:MNS-RNS} is $\ToverW=0.0419$, 
and that of the maximum mass model with maximal rotation (blue filled circle in 
the top panel of Fig.~\ref{fig:M_prhoc}) is around $\ToverW=0.050$, 
while for the maximum mass model of EMV$^-$IV (the one in Fig.~\ref{fig:MNS-RNS}), 
the ratio value becomes $\MoverW =0.11$.  

In Fig.~\ref{fig:M_prhoc_hyper}, we show the $\Madm$--$\prhoc$ plots (top panel) and 
$\MoverW$--$\prhoc$ plots (bottom panel) for EMV$^+$ and FF models.  
For EMV$^+$ models except for $\prhoc=0.3$, the parameter $\Lambda_1$ is increased from 
0.4--0.55 every 0.05 from bottom to top for each $\prhoc$.  In this sequence, it is found that 
the mass of the solution (and the magnetic fields accordingly) increases rapidly at a certain value of 
$\Lambda_1$.  Therefore, at some point, we fix the deformation parameter $R_z/R_0$ (and $\prhoc$), 
then gradually increase the value of $\Lambda_1$.  The largest $\Lambda_1$ for each $\prhoc$ is 
$\Lambda_1=0.567$, 0.532, 0.491, and 0.432 for $\prhoc=0.1527579$, 0.1733394, 0.2021665, 0.2558203 
($M_0=1.35$, 1.4, 1.45, and 1.5), respectively, where the degree of rotation $\ToverW$ becomes 
the smallest, simultaneously.  
For $\prhoc=0.3$, plotted are $\Lambda_1=0.3$, 0.35, 0.399 from bottom to top using diamonds.  
The EMV$^+$ models with the largest $\Lambda_1$ at these $\prhoc$ ($M_0=1.35$, 1.4, 1.45, 1.5 
and $\prhoc=0.3$) are the hypermassive solutions.  

The caveat is that, for each $\prhoc$, it may be possible to compute a heavier solution associated 
with stronger magnetic fields for a larger deformation (a smaller value of $R_z/R_0$).  Another caveat is 
that it may also be possible to compute a heavier (and eventually hypermassive) solution for 
EMV$^-$ models by applying a similar procedure to change the magnetic field parameters gradually.

For the FF model, a sequence of solutions is obtained systematically from the lower to higher 
$\prhoc$, as plotted also in Fig.~\ref{fig:M_prhoc_hyper}, which are hypermassive except for 
the lowest $\prhoc$ model (with $M_0=1.25$).  It is noted that the values of $\MoverW$ of 
these magnetically hypermassive solutions are much larger than $2\ToverW$ of the maximum mass 
model of a rotating star (the blue dot model in the top panel), which is $2\ToverW \approx 0.099$.

\section{Discussion}
\label{sec:Discussion}

In the present computation for EMV$^-$I--IV in Fig.~\ref{fig:M_prhoc}, 
the excess of the mass of relativistic compact stars only due to the magnetic 
fields is around $\sim 12\%$ of the maximum mass of the TOV star.  
This is comparable but smaller than the excess of the mass due to 
uniform rotation, which is $\sim 15\%$ for the $\Gamma=2$ EOS.  
It is not clear that the selected parameter set is close to the limit 
of the strongest magnetic field of such model - a compact star surrounded by 
electromagnetic vacuum and associated with a force-free tunnel.  
There is a possibility for computing hypermassive models having stronger magnetic 
fields than the EMV$^-$IV model.  It is also possible that hypermassive solutions 
exist with $\Lambda_0 < 0$ associated with a force-free magnetosphere.  
The reason for not having proceeded to a stronger magnetic field than the EMV$^-$IV 
model is solely a lack of computing resources, which may be resolved by the time 
a parallel version of the code is available.

It is known that various (non-magnetized) differentially rotating solutions 
become hypermassive, as high as $\sim 20-100\%$, depending on the differential 
rotation models \cite{RNSdiff2}.  
We found that models with $\Lambda_0 > 0$, EMV$^+$ and FF models, can be 
magnetically hypermassive, and the star becomes more massive, especially when 
the toroidal region of strong magnetic field near the equatorial surface become wider, 
as in the FF model.

As discussed in Sec.~\ref{sec:comparison}, 
although the shape of the magnetized neutron star resembles that of a rotating one, 
its gravitational field, such as the magnitude of shift $\tbeta_y$, is noticeably different. 
Therefore for a given shape of a NS, we expect that the effect of frame-dragging will 
be weaker in a strongly magnetized neutron star, and it may be used as a probe of its 
magnetic field.  In addition, by observing the X-rays emitted from the neutron star's 
surface, the X-ray telescope NICER \cite{NICER} was able to constrain the neutron star 
mass-radius relation and the EOS of dense matter.  Photons emitted from the star's surface 
reach an observer by following curved paths that characterize the neutron star's gravitational 
field, which also affects their energy.  Similar methodology is also used by the Chandra and 
XMM-Newton telescopes.  Therefore, although a magnetar and a rotating neutron star have 
similar shapes, their different gravitational fields will lead to different X-ray emissions, 
which in principle may also be used to probe a strong magnetic field.

It has been discussed that a large deformation of a compact star due to strong magnetic fields 
may be detectable if the star's magnetic axis is tilted with respect to the rotation axis.  
The magnetic equilibriums, including those presented in this paper, may approximate such 
solutions if the tilted angle is small and the rotation is slow.  
Although a detection of continuous gravitational waves emitted from the wobbling of such a 
star is quite challenging because of their small amplitudes, recent advances in gravitational 
wave data analysis may enable their detection in the near future \cite{LIGOScientific:2025kei}. 
It may also be promising to probe the deformation from a detection of free precession 
observed as phase modulations of X-ray pulsations, as reported in \cite{Makishima:2014dua}

In \cite{MRNSsimulationsUIUC}, we have performed preliminary full GRMHD simulations of 
such highly magnetized rotating compact stars starting from a selected set of 
equilibrium solutions obtained from the \cocal\ code used in the present work.  
We have found that stability of such highly magnetized rotating compact stars depends 
strongly on the magnetic field configurations, such as the mixture of the toroidal 
and poloidal components, with some models being highly unstable while others are 
stable over many Alfv\'en timescales.  Stability of magnetized compact stars
is challenging to characterize due to the large parameter space spanned by magnetized 
equilibria (which includes density, rotation, magnetic field configuration and strength)
(see, e.g.~\cite{Shapiro:2000zh,MNS_instabilities}).  In Appendix \ref{sec:App_stability}, 
we also discuss the turning-point method, which gives a sufficient condition for instability.  
Further simulations including current magnetically supramassive and hypermassive 
models as initial data, will shed more light on the properties
of these extraordinary compact objects.

The other extension of the present work is to take into account realistic EOS.  
The extremely strong magnetic field in our model exist not only at the core 
of the compact star but also concentrated near the equatorial surface.  Therefore, it is 
particularly important to take into account the interplay between such extreme magnetic 
fields and the low density region especially the crust.  For example, in our models with 
an FF tunnel, the wall of the FF tunnel should be a crust in reality.  Such computation 
may be possible with a recent update of the tabulated piecewise polytropic EOS 
\cite{Suleiman:2022egw}.

\begin{acknowledgments}
This work was supported by 
JSPS Grant-in-Aid for Scientific Research(C) 22K03636, 18K03624, 25K07274, 18K03606, 
24K07053, 21K03556, 20H04728, and NSF Grants No.PHY-2308242 and No.OAC-2310548 
to the University of Illinois Urbana-Champaign, 
and Marie Sklodowska-Curie grant agreement No.753115.  
AT acknowledges support from the National Center
for Supercomputing Applications (NCSA) at the University of Illinois at
Urbana-Champaign through the NCSA Fellows program.
%
%
%
\end{acknowledgments}

\appendix

\section{Physical quantities of selected solutions}
\label{sec:App_physq}

In Tables \ref{tab:MRNS_EMV-IV}--\ref{tab:MRNS_FF_EMF}, physical quantities of selected solutions for 
EMV$^-$, EMV$^+$, and FF models are presented.  Those selected solutions are the EMV$^-$IV model 
in Tables \ref{tab:MRNS_EMV-IV} and \ref{tab:MRNS_EMV-IV_EMF}, 
hypermassive solutions of EMV$^+$ in Tables \ref{tab:MRNS_EMV+} and \ref{tab:MRNS_EMV+_EMF}, 
and solutions of the FF model in Tables \ref{tab:MRNS_FF} and \ref{tab:MRNS_FF_EMF}.  
Definitions of these physical quantities are detailed in Appendix F of Paper I.  
As explained in Sec.\ref{sec:non-rotatingMRNS}, 
our solutions are approximately non-rotating solutions, which are obtained from 
our magnetized rotating star code by setting the rotation as small as possible 
(in practice, maximizing the axis ratio $R_z/R_0$ under a fixed $\prhoc$ for each solution).  
In the tables, we also show the rotation related quantities for reference, which are 
the angular velocity $\Omegac$, the normalized angular momentum $J/\Madm^2$, and $\ToverW$.  
In the virial relation $\Ivir=2{\cal T} +3\Pi + {\cal M} + {\cal W}=0 $, the 
contribution of the kinetic term $2{\cal T}$ is less than 5\% (typically around 
1\%) of the magnetic term ${\cal M}$ and less than 0.4\% of the internal energy term 
$3\Pi$ for the EMV$^-$ model.  
For the FF model, the contribution of the kinetic term $2{\cal T}$ is less than 10\% 
of the magnetic term ${\cal M}$, and less than 2\% of the internal energy term $3\Pi$.  

As mentioned in Sec.~\ref{sec:SMHMsequence}, for the hypermassive solutions of EMV$^+$ models, 
we further adjust the magnetic field parameter $\Lambda_1$ to minimize the degree of rotation.  
Then the contribution of kinetic term $2{\cal T}$ is less than 1.2\%
of the magnetic term ${\cal M}$, and less than  0.2\% of the internal energy term $3\Pi$.

\begin{table*}
\caption{Physical quantities of EMV$^-$IV model solutions.  
Listed quantities are $\prhoc$, $\rhoc$, 
the equatorial radius in proper length $\bar{R}_0$, 
the ratio of the polar to equatorial radii in the in proper length $\bar{R}_z/\bar{R}_0$, 
the ADM mass $\Madm$, the rest mass $M_0$, the proper mass $\MP$, 
the angular velocity $\Omegac$, the angular momentum $J$, 
the ratio of the kinetic energy to the gravitational energy $\ToverW$, 
and a residual of the equality of the Komar mass $\MK$ and the ADM mass $\Madm$.  
Units of dimensional quantities are in $G=c=\Msol=1$ units except for $\rhoc$ in [$10^{15}\mbox{g/cm}^3$].  
To convert a unit of length from $G=c=\Msol=1$ to [km], multiply 
$G\Msol/c^2=1.477$[km] (see, Table \ref{tab:TOV_solutions}).
Details of the definitions for these quantities are found in Appendix F of Paper I.  
}
\label{tab:MRNS_EMV-IV}
\begin{tabular}{cccccccclcc}
\hline
$\prhoc$ & $\rhoc$ & 
$\bar{R}_0$ & $\bar{R}_z/\bar{R}_0$ & 
$\Madm$ & $M_0$ & 
$\MP$ & 
$\Omegac$ & 
$J/\Madm^2$ & 
$\ToverW$ & 
$|1-\MK/\Madm|$ 
\\
\hline
0.1232211 & 1.0717 & 10.089 & 0.80475 & 1.2731 & 1.3622 & 1.4434 & $3.7872 \times 10^{-3}$ & 0.073333 & $8.2649 \times 10^{-4}$ & $1.2125 \times 10^{-5}$ \\
0.1365875 & 1.1880 & 10.088 & 0.78812 & 1.3289 & 1.4254 & 1.5188 & $5.9551 \times 10^{-3}$ & 0.10909  & $1.7934 \times 10^{-3}$ & $4.2166 \times 10^{-5}$ \\
0.1527579 & 1.3286 & 10.019 & 0.77776 & 1.3784 & 1.4815 & 1.5892 & $5.2450 \times 10^{-3}$ & 0.090385 & $1.1968 \times 10^{-3}$ & $8.0411 \times 10^{-5}$ \\
0.1733394 & 1.5076 & 9.9689 & 0.76201 & 1.4317 & 1.5415 & 1.6673 & $6.7504 \times 10^{-3}$ & 0.10910  & $1.6732 \times 10^{-3}$ & $1.0440 \times 10^{-4}$ \\
0.2021665 & 1.7583 & 9.8665 & 0.74720 & 1.4817 & 1.5964 & 1.7460 & $6.5455 \times 10^{-3}$ & 0.098159 & $1.2695 \times 10^{-3}$ & $2.0349 \times 10^{-4}$ \\
0.2558203 & 2.2250 & 9.6580 & 0.72899 & 1.5292 & 1.6439 & 1.8337 & $4.2548 \times 10^{-3}$ & 0.057361 & $3.8256 \times 10^{-4}$ & $4.7777 \times 10^{-4}$ \\
0.3000000 & 2.6092 & 9.4690 & 0.72123 & 1.5387 & 1.6493 & 1.8680 & $3.7189 \times 10^{-3}$ & 0.046642 & $2.3193 \times 10^{-4}$ & $5.5051 \times 10^{-4}$ \\
0.3500000 & 3.0441 & 9.2852 & 0.71365 & 1.5351 & 1.6380 & 1.8854 & $6.4798 \times 10^{-3}$ & 0.076031 & $5.6400 \times 10^{-4}$ & $7.7099 \times 10^{-4}$ \\
0.4000000 & 3.4790 & 9.0859 & 0.71131 & 1.5180 & 1.6126 & 1.8848 & $8.9551 \times 10^{-3}$ & 0.098861 & $8.9010 \times 10^{-4}$ & $8.3915 \times 10^{-4}$ \\
\hline
\end{tabular}
\end{table*}
\begin{table*}
\caption{Continued from Table \ref{tab:MRNS_EMV-IV}, 
listed quantities are the maximum values 
of poloidal and toroidal magnetic fields, $\Bpolmax$ and $\Btormax$, 
the ratios of poloidal and toroidal magnetic field energies, 
and electric field energy, to the total electromagnetic field energy,
${\cal M}_{\rm pol}/{\cal M}$, ${\cal M}_{\rm tor}/{\cal M}$, and 
${\cal M}_{\rm ele}/{\cal M}$, respectively, 
the ratios of the internal, and electromagnetic field energies 
to the gravitational energy, $\PioverW$, and $\MoverW$, respectively, 
the virial constant divided by the gravitational energy, 
$\IviroverW$, and the electric charge contribution from the volume integral 
of the star $Q_M$.  
Details of the definitions are found in Appendix F in Paper I.  
The maximums of magnetic field components $\Bpolmax$ and $\Btormax$ are 
defined by those of spatial Faraday tensor $F_{ab}$ in Cartesian coordinates, 
$\Bpol:=F_{xy}$ and $\Btor:=-F_{xz}$.  Units of $\Bpolmax$ and $\Btormax$ are in [G], 
and $Q_M$ is in $G=c=\Msol=4\pi\epsilon_0=1$ units.  
}
\label{tab:MRNS_EMV-IV_EMF}
\begin{tabular}{cccclclcc}
\hline
$\Bpolmax$[G] & $\Btormax$[G] & 
${\cal M}_{\rm pol}/{\cal M}$ & ${\cal M}_{\rm tor}/{\cal M}$ &\ ${\cal M}_{\rm ele}/{\cal M}$ & 
$\PioverW$ & $\ \MoverW$ & $\IviroverW$ & $Q_M$ 
\\
\hline
$1.1140\times 10^{18}$ & $1.7355\times 10^{18}$ & 0.95189 & 0.047613 & 0.00049664 & 0.30964 & 0.069767 & $3.4254\times 10^{-4}$ & $8.3087\times 10^{-3}$ \\
$1.3320\times 10^{18}$ & $2.0905\times 10^{18}$ & 0.94962 & 0.049140 & 0.0012389  & 0.30687 & 0.076268 & $4.6307\times 10^{-4}$ & $1.4851\times 10^{-2}$ \\
$1.6388\times 10^{18}$ & $2.5257\times 10^{18}$ & 0.94837 & 0.050676 & 0.00095703 & 0.30504 & 0.083080 & $5.9062\times 10^{-4}$ & $1.4621\times 10^{-2}$ \\
$2.0122\times 10^{18}$ & $3.1083\times 10^{18}$ & 0.94670 & 0.051733 & 0.0015697  & 0.30245 & 0.089946 & $6.3304\times 10^{-4}$ & $2.1214\times 10^{-2}$ \\
$2.5966\times 10^{18}$ & $3.8940\times 10^{18}$ & 0.94539 & 0.053168 & 0.0014468  & 0.30003 & 0.098263 & $8.7784\times 10^{-4}$ & $2.3309\times 10^{-2}$ \\
$3.8449\times 10^{18}$ & $5.4580\times 10^{18}$ & 0.94443 & 0.054984 & 0.00058141 & 0.29728 & 0.10881  & $1.4214\times 10^{-3}$ & $1.7599\times 10^{-2}$ \\
$4.9521\times 10^{18}$ & $6.4260\times 10^{18}$ & 0.94448 & 0.055104 & 0.00041987 & 0.29690 & 0.11026  & $1.4337\times 10^{-3}$ & $1.6139\times 10^{-2}$ \\
$6.2841\times 10^{18}$ & $7.5077\times 10^{18}$ & 0.94377 & 0.055026 & 0.0012012  & 0.29691 & 0.10987  & $1.7143\times 10^{-3}$ & $2.8945\times 10^{-2}$ \\
$7.6487\times 10^{18}$ & $8.2906\times 10^{18}$ & 0.94351 & 0.054341 & 0.0021485  & 0.29802 & 0.10585  & $1.7019\times 10^{-3}$ & $3.9487\times 10^{-2}$ \\
%
%
%
\hline
\end{tabular}
\end{table*}
\begin{table*}
\caption{Same as Table \ref{tab:MRNS_EMV-IV}, but for hypermassive solutions of the EMV$^+$ model.  
}
\label{tab:MRNS_EMV+}
\begin{tabular}{cccccccclcc}
\hline
$\prhoc$ & $\rhoc$ & 
$\bar{R}_0$ &  $\bar{R}_z/\bar{R}_0$ & 
$\Madm$ & $M_0$ & 
$\MP$ & 
$\Omegac$ & 
$J/\Madm^2$ & 
$\ToverW$ & 
$|1-\MK/\Madm|$ 
\\
\hline
0.1527579 & 1.3286 & 10.180 & 0.65303 & 1.5988 & 1.6982 & 1.8338 & $4.1448 \times 10^{-3}$ & $7.5050 \times 10^{-2}$ & $6.8280 \times 10^{-4}$ & $9.6214 \times 10^{-5}$ \\
0.1733394 & 1.5076 & 9.9312 & 0.66690 & 1.6149 & 1.7157 & 1.8678 & $4.8454 \times 10^{-3}$ & $8.1499 \times 10^{-2}$ & $7.8254 \times 10^{-4}$ & $1.2245 \times 10^{-4}$ \\
0.2021665 & 1.7583 & 9.5720 & 0.69319 & 1.6081 & 1.7101 & 1.8802 & $5.0466 \times 10^{-3}$ & $7.7271 \times 10^{-2}$ & $6.8727 \times 10^{-4}$ & $1.2398 \times 10^{-4}$ \\
0.2558203 & 2.2250 & 9.0977 & 0.72083 & 1.6031 & 1.7015 & 1.9056 & $7.0195 \times 10^{-3}$ & $9.4288 \times 10^{-2}$ & $9.5463 \times 10^{-4}$ & $1.3215 \times 10^{-4}$ \\
0.3000000 & 2.6092 & 8.8137 & 0.72963 & 1.6016 & 1.6903 & 1.9210 & $6.6872 \times 10^{-3}$ & $8.3190 \times 10^{-2}$ & $6.8360 \times 10^{-4}$ & $1.5465 \times 10^{-4}$ \\
\hline
\end{tabular}
\end{table*}
\begin{table*}
\caption{Same as Table \ref{tab:MRNS_EMV-IV_EMF}, but continued from 
Table \ref{tab:MRNS_EMV+} for hypermassive solutions of the EMV$^+$ model.  
}
\label{tab:MRNS_EMV+_EMF}
\begin{tabular}{cccclclcc}
\hline
$\Bpolmax$[G] & $\Btormax$[G] & 
${\cal M}_{\rm pol}/{\cal M}$ & ${\cal M}_{\rm tor}/{\cal M}$ &\ ${\cal M}_{\rm ele}/{\cal M}$ & 
$\PioverW$ & $\ \MoverW$ & $\IviroverW$ & $Q_M$ 
\\
\hline
$2.9720 \times 10^{18}$ & $2.8049 \times 10^{18}$ & 0.93698 & 0.061975 & 0.0010454  & 0.27145 & 0.18424 & $4.7484 \times 10^{-5}$ & $1.9958 \times 10^{-2}$ \\
$3.4119 \times 10^{18}$ & $3.3710 \times 10^{18}$ & 0.93272 & 0.066697 & 0.00058064 & 0.27235 & 0.18129 & $1.1263 \times 10^{-4}$ & $2.3040 \times 10^{-2}$ \\
$3.9288 \times 10^{18}$ & $4.0031 \times 10^{18}$ & 0.92819 & 0.071213 & 0.00059546 & 0.27645 & 0.16920 & $7.7621 \times 10^{-5}$ & $2.2398 \times 10^{-2}$ \\
$5.0105 \times 10^{18}$ & $5.3841 \times 10^{18}$ & 0.92026 & 0.078673 & 0.0010673  & 0.27973 & 0.15885 & $3.7771 \times 10^{-5}$ & $2.9312 \times 10^{-2}$ \\
$6.1059 \times 10^{18}$ & $6.8819 \times 10^{18}$ & 0.91487 & 0.084212 & 0.00092024 & 0.27919 & 0.16102 & $3.9331 \times 10^{-5}$ & $2.7938 \times 10^{-2}$ \\
%
%
%
\hline
\end{tabular}
\end{table*}
\begin{table*}
\caption{Same as Table \ref{tab:MRNS_EMV-IV}, but for solutions of the FF model.  }
\label{tab:MRNS_FF}
\begin{tabular}{cccccccclcc}
\hline
$\prhoc$ & $\rhoc$ & 
$\bar{R}_0$ &  $\bar{R}_z/\bar{R}_0$ & 
$\Madm$ & $M_0$ & 
$\MP$ & 
$\Omegac$ & 
$J/\Madm^2$ & 
$\ToverW$ & 
$|1-\MK/\Madm|$ 
\\
\hline
0.1232211 & 1.0717 & 11.797 & 0.64346 & 1.5242 & 1.6120 & 1.7032 & $4.6195 \times 10^{-3}$ & 0.10147 & $1.4896 \times 10^{-3}$ & $1.8156 \times 10^{-4}$ \\
0.1365875 & 1.1880 & 11.874 & 0.62751 & 1.6073 & 1.6997 & 1.8047 & $4.7697 \times 10^{-3}$ & 0.10075 & $1.4260 \times 10^{-3}$ & $2.0718 \times 10^{-4}$ \\
0.1527579 & 1.3286 & 11.949 & 0.61215 & 1.6898 & 1.7854 & 1.9065 & $5.1500 \times 10^{-3}$ & 0.10443 & $1.4784 \times 10^{-3}$ & $2.4385 \times 10^{-4}$ \\
0.1733394 & 1.5076 & 12.030 & 0.59778 & 1.7638 & 1.8628 & 2.0026 & $7.8792 \times 10^{-3}$ & 0.15225 & $3.0411 \times 10^{-3}$ & $3.9391 \times 10^{-4}$ \\
0.2021665 & 1.7583 & 11.977 & 0.59057 & 1.8104 & 1.9125 & 2.0754 & $1.0819 \times 10^{-2}$ & 0.19418 & $4.8158 \times 10^{-3}$ & $6.0221 \times 10^{-4}$ \\
0.2558203 & 2.2250 & 11.657 & 0.59228 & 1.8132 & 1.9178 & 2.1172 & $1.5133 \times 10^{-2}$ & 0.23687 & $6.9993 \times 10^{-3}$ & $9.5989 \times 10^{-4}$ \\
0.3000000 & 2.6092 & 11.354 & 0.59769 & 1.7806 & 1.8861 & 2.1114 & $1.8779 \times 10^{-2}$ & 0.26457 & $8.7151 \times 10^{-3}$ & $1.3012 \times 10^{-3}$ \\
0.3500000 & 3.0441 & 10.910 & 0.60803 & 1.7521 & 1.8444 & 2.0968 & $1.5788 \times 10^{-2}$ & 0.20458 & $4.9043 \times 10^{-3}$ & $9.2890 \times 10^{-4}$ \\
0.4000000 & 3.4790 & 10.521 & 0.61762 & 1.7199 & 1.7969 & 2.0730 & $1.0074 \times 10^{-2}$ & 0.12196 & $1.6377 \times 10^{-3}$ & $5.1568 \times 10^{-4}$ \\
\hline
\end{tabular}
\end{table*}
\begin{table*}
\caption{Same as Table \ref{tab:MRNS_EMV-IV_EMF}, but continued from 
Table \ref{tab:MRNS_FF} for solutions of the FF model.}
\label{tab:MRNS_FF_EMF}
\begin{tabular}{cccclclcc}
\hline
$\Bpolmax$[G] & $\Btormax$[G] & 
${\cal M}_{\rm pol}/{\cal M}$ & ${\cal M}_{\rm tor}/{\cal M}$ &\ ${\cal M}_{\rm ele}/{\cal M}$ & 
$\PioverW$ & $\ \MoverW$ & $\IviroverW$ & $Q_M$ 
\\
\hline
$1.3012 \times 10^{18}$ & $8.7929 \times 10^{17}$ & 0.91363 & 0.085283 & 0.0010845 & 0.26647 & 0.19707 & $5.5045 \times 10^{-4}$ & $2.6058 \times 10^{-2}$ \\
$1.5460 \times 10^{18}$ & $1.0570 \times 10^{18}$ & 0.91059 & 0.088250 & 0.0011646 & 0.26163 & 0.21169 & $5.7343 \times 10^{-4}$ & $3.0924 \times 10^{-2}$ \\
$1.8508 \times 10^{18}$ & $1.2686 \times 10^{18}$ & 0.90733 & 0.091307 & 0.0013654 & 0.25716 & 0.22495 & $6.1887 \times 10^{-4}$ & $3.8143 \times 10^{-2}$ \\
$2.2125 \times 10^{18}$ & $1.4796 \times 10^{18}$ & 0.90285 & 0.093935 & 0.0032145 & 0.25429 & 0.23003 & $1.0062 \times 10^{-3}$ & $6.5191 \times 10^{-2}$ \\
$2.6943 \times 10^{18}$ & $1.7013 \times 10^{18}$ & 0.89718 & 0.096880 & 0.0059420 & 0.25495 & 0.22408 & $1.4439 \times 10^{-3}$ & $9.4799 \times 10^{-2}$ \\
$3.5689 \times 10^{18}$ & $1.9747 \times 10^{18}$ & 0.88814 & 0.101067 & 0.0107966 & 0.26130 & 0.20011 & $2.0017 \times 10^{-3}$ & $1.2954 \times 10^{-1}$ \\
$4.2511 \times 10^{18}$ & $2.0837 \times 10^{18}$ & 0.88139 & 0.103089 & 0.0155256 & 0.26801 & 0.17606 & $2.4641 \times 10^{-3}$ & $1.4907 \times 10^{-1}$ \\
$5.3713 \times 10^{18}$ & $2.4954 \times 10^{18}$ & 0.88086 & 0.109152 & 0.0099903 & 0.27096 & 0.17588 & $1.4309 \times 10^{-3}$ & $1.2172 \times 10^{-1}$ \\
$6.5857 \times 10^{18}$ & $2.9191 \times 10^{18}$ & 0.88142 & 0.114841 & 0.0037342 & 0.27322 & 0.17652 & $5.4074 \times 10^{-4}$ & $7.5347 \times 10^{-2}$ \\
%
%
%
\hline
\end{tabular}
\end{table*}

\section{Discussion on the stability of non-rotating magnetized relativistic stars}
\label{sec:App_stability}

Stability of magnetized stars with or without rotation is complicated.  
Even weak magnetic fields may destabilize the star depending on the configurations 
of magnetic fields, and a rotation may sometimes stabilize it \cite{Kiuchi:2008ss}.  
Suppose that a stable magnetic configuration and a strength is found for a non-rotating 
magnetized model, it is natural to expect that the solutions with the lower central 
density may be dynamically stable against radial perturbation, similarly to TOV 
solutions.  For the uniformly rotating (non-magnetized) relativistic stars, 
the gravitational mass vs the central density, $M(\rhoc)$, plot along a constant 
rest mass sequence can be used to identify the dynamical stability of the solutions.  
A turning point theorem states that the stability changes at a turning point of 
$M(\rhoc)$ along a sequence of constant rest mass, which is similar to a 
change of stability at the maximum mass of the TOV solutions \cite{RNS}.  
It appears, however, that the turning point method cannot be applied for 
the data set of non-rotating magnetized stars we computed so far.  
Plots of $M(\rhoc)$ for constant rest mass sequences obtained from 
interpolating the systematic data used in Fig.~\ref{fig:M_prhoc} was nearly 
flat - there was no turning point along the constant rest mass sequence 
for the data set.  

One of the conditions to apply the turning point method to non-magnetized rotating 
relativistic stars is to collect a set of solutions that satisfies the first law 
relation 
\beq
\delta M = \Omega\, \delta J
\label{eq:1stlaw}
\eeq
for nearby equilibrium solutions, where 
$J$ is the total angular momentum.  It is 
known to be satisfied for the constant rest mass and entropy sequences of the uniformly 
rotating solutions, and hence used as stability criteria \cite{Friedman:1988er,CST,CST94}. 
\!\!\!\footnote{It is also suggested that the turning point of $M(\rhoc)$ is close to the 
point of the onset of instability for differentially rotating solutions, although 
the first law is not strictly satisfied \cite{Weih:2017mcw}.}
As derived by Carter in \cite{Carter73}, 
\!\!\!\footnote{We use the notation used in 
our paper \cite{Uryu:2010su}, replacing a helical Killing vector $k^\alpha$ with a timelike 
Killing vector $t^\alpha$, then discard the terms related to the black holes.} 
the mass variation formula for the magnetized self-gravitating matter becomes, 
\beqn
\delta M  
&=& 
\int_\Sigma \left\{\,\frac{T}{u^t} \Dl (s \, \rho u^\alpha\, \dSa)
\right.
\,+\,\frac{h-Ts}{u^t}\Dl(\rho u^\alpha\, \dSa)
\nonumber \\ 
&&\quad
\,+\,v^\beta \Dl(hu_\beta\,\rho u^\alpha\,\dSa) 
\,-\, A_\beta t^\beta \, \Dl (j^\alpha \dSa)
\nonumber \\ 
&&\left.\phantom{\frac12}
\,-\,(j^\alpha t^\beta - j^\beta t^\alpha )
\Dl A_\beta \,\dSa
\,\right\}, 
\label{eq:1stlaw_org}
\eeqn
where $T$ is the temperature, $s$ the entropy per baryon, and 
$h$ the relativistic enthalpy.  The operator $\Dl$ denotes the Lagrangian 
perturbation that map a congruence of the fluid's world line to the perturbed 
congruence.  For the uniform rotation, $v^\alpha = \Omega \phi^\alpha$, 
the third term of the integrand of the RHS of Eq.~(\ref{eq:1stlaw_org}) becomes 
$\Omega\,\dl J$.  Since the perturbations of the isentropic flow and the rest 
mass conservation law are, respectively, $\Dl s = 0$, and $\Dl(\rho u^\alpha\,\dSa)=0$, 
the first two terms of the integrand are eliminated.  Therefore, for the 
non-magnetized uniformly rotating stars, the relation (\ref{eq:1stlaw}) is satisfied 
along the rest mass constant and thermal equilibriums.  

For the magnetized rotating relativistic stars, a relation as simple as 
Eq.~(\ref{eq:1stlaw}) is not known.  As discussed in \cite{Uryu:2010su}, 
the ideal MHD condition $\Fabd u^\alpha=0$ implies the conservation of magnetic flux 
$\Lie_u \Fabd = 0$, and its perturbed version $\Delta \Fabd = 0$ 
is derived when the perturbed state also satisfies the ideal MHD condition.  
Since $\Delta \Fabd = d(\Dl A)_\albe=0$, the Poincar\'e lemma guarantees the 
existence of a scalar field $\Psi$ that satisfies a relation 
$\Dl A_\alpha=\na_\alpha \Psi$.  Then, substituting the relation, and integrating 
by part, the fifth term in the integrand of Eq.~(\ref{eq:1stlaw_org}) vanishes
\cite{Uryu:2010su} when the surface current vanishes or is static.  
For the non-rotating magnetized case, $u^\alpha = u^t t^\alpha$, the ideal MHD 
condition implies $A_\alpha t^\alpha = $constant.  Then, with a gauge condition 
$\Dl t^\alpha = 0$ \cite{Friedman:2001pf}, the fourth term becomes 
\beq
\int_\Sigma A_\beta t^\beta \, \Dl (j^\alpha \dSa)
\,=\,A_t\,\dl \QE
\label{eq:1stlaw_EMterms}
\eeq
where $\dl\QE$ is a variation of the electric charge.  
Hence, if in addition the electric charge is conserved, $\delta Q^{\rm E} = 0$, and 
the magnetic flux is conserved (as in ideal MHD), then, 
\beq
\dl M = 0
\label{eq:1stlaw_dlM}
\eeq
is satisfied. 
\!\!\!\footnote{For the other example of the current that eliminates 
the terms in (\ref{eq:1stlaw_EMterms}), see \cite{Uryu:2010su}.  }

It should be noted that such a simple relation ({\ref{eq:1stlaw_dlM}) appears 
to be satisfied for the non-rotating magnetized relativistic stars in equilibrium 
under the ideal MHD.  
It seems, however, the first law alone is not enough to apply the turning 
point method to determine the stability of those stars.  
As in the case of differentially rotating stars, there are a variety of 
magnetic configurations, as in obvious examples like purely poloidal, purely toroidal, 
or mixed poloidal and toroidal magnetic fields.  
Presumably, to apply the turning point method, it is necessary to prepare 
a data set whose member solutions have closely related magnetic configurations, 
at least, conserved magnetic flux, and possibly the magnetic helicity.  


\end{document}